\documentclass[pra,twocolumn,superscriptaddress,showpacs,floatfix,longbibliography]{revtex4-1}
\usepackage{mathrsfs}
\usepackage{amssymb, amsbsy, amsmath, latexsym, dsfont, array, layout,
graphicx,mathrsfs,color,ulem,bm}

\newcommand{\one}{\mathds{1}}
\newcommand{\ket}[1]{\left|{#1}\right\rangle}
\newcommand{\bra}[1]{\left\langle{#1}\right|}

\begin{document}

\title{Higher winding number in a non-unitary photonic quantum walk}
\author{Lei Xiao}
\affiliation{Beijing Computational Science Research Center, Beijing 100084, China}
\affiliation{Department of Physics, Southeast University, Nanjing 211189, China}
\author{Xingze Qiu}
\affiliation{Key Laboratory of Quantum Information, University of Science and Technology of China, CAS, Hefei 230026, China}
\affiliation{Synergetic Innovation Center in Quantum Information and Quantum Physics, University of Science and Technology of China, CAS, Hefei 230026, China}
\author{Kunkun Wang}
\affiliation{Beijing Computational Science Research Center, Beijing 100084, China}
\affiliation{Department of Physics, Southeast University, Nanjing 211189, China}
\author{Zhihao Bian}
\affiliation{Beijing Computational Science Research Center, Beijing 100084, China}
\affiliation{Department of Physics, Southeast University, Nanjing 211189, China}
\author{Xiang Zhan}
\affiliation{Beijing Computational Science Research Center, Beijing 100084, China}
\affiliation{Department of Physics, Southeast University, Nanjing 211189, China}
\author{Hideaki Obuse}
\affiliation{Department of Applied Physics, Hokkaido University, Sapporo 060-8628, Japan}
\author{Barry C. Sanders}
\affiliation{Hefei National Laboratory for Physical Sciences at Microscale, University of Science and Technology of China, CAS, Hefei 230026, China}
\affiliation{Institute for Quantum Science and Technology, University of Calgary, Alberta T2N 1N4, Canada}
\affiliation{Program in Quantum Information Science, Canadian Institute for Advanced Research, Toronto, Ontario M5G 1M1, Canada}
\author{Wei Yi}\email{wyiz@ustc.edu.cn}
\affiliation{Key Laboratory of Quantum Information, University of Science and Technology of China, CAS, Hefei 230026, China}
\affiliation{Synergetic Innovation Center in Quantum Information and Quantum Physics, University of Science and Technology of China, CAS, Hefei 230026, China}
\author{Peng Xue}\email{gnep.eux@gmail.com}
\affiliation{Beijing Computational Science Research Center, Beijing 100084, China}
\affiliation{Department of Physics, Southeast University, Nanjing 211189, China}
\affiliation{State Key Laboratory of Precision Spectroscopy, East China Normal University, Shanghai 200062, China}

\begin{abstract}
Topological matter exhibits exotic properties yet phases characterized by large topological invariants are difficult to implement, despite rapid experimental progress. A promising route toward higher topological invariants is via engineered Floquet systems, particularly in photonics, where flexible control holds the potential of extending the study of conventional topological matter to novel regimes.
Here we implement a one-dimensional photonic quantum walk to explore large winding numbers. By introducing partial measurements and hence loss into the system, we detect winding numbers of three and four in multi-step non-unitary quantum walks, which agree well with theoretical predictions. Moreover, by probing statistical moments of the walker, we identify locations of topological phase transitions in the system, and reveal the breaking of pseudo-unitary near topological phase boundaries. As the winding numbers are associated with non-unitary time evolution, our investigation enriches understanding of topological phenomena in non-unitary settings.
\end{abstract}


\maketitle

Topological phases are typically characterized by integer-valued topological invariants, associated with the emergence of robust edge states through the so-called bulk-boundary correspondence~\cite{HKrmp10,QZrmp11,Ryu10,Kane10}. Recent experiments reveal and characterize topological edge states and bulk topological invariants in settings ranging from condensed matter~\cite{HKrmp10,CYL09,Xia09} to synthetic systems~\cite{WCJS09,LJS14,KB+12,LL15, Cardano2016,PTsymm2,PBKMS15,BKMM13,Chong15,Hubers,Khanikaevnc,ETHcoldatom14,Weitz16,Gadway16,Bloch13,CBG15,Weitenberg2016,RFR+17,FRH+16,pxprl,Cardano2017,silberhorn16,Zeunerprl}. However, the experimentally detected topological invariants are typically small and limited to two~\cite{Bloch13,CBG15,Weitenberg2016,RFR+17,pxprl,Cardano2017,silberhorn16,Zeunerprl,FRH+16}. Whereas bands with Chern numbers greater than two have been engineered in photonic materials in two dimensions~\cite{LL15}, direct detection of Chern numbers greater than two has yet to be achieved.
In one dimension, while topological phases with large winding numbers have been theoretically studied, e.g., in quantum transport~\cite{RLL16} or in quantum-walk dynamics~\cite{AO13,KMKO16}, experimental realization is still lacking. Realizing systems with large topological invariants, whether large Chern numbers in two dimensions~\cite{LL15, Chern1,Chern2,Chern3} or large winding numbers in one dimension~\cite{RLL16, AO13,KMKO16}, is fundamentally important goal for the study of topological matter.

A promising platform for detection of large bulk topological invariants is synthetic Floquet topological systems, where winding numbers of two have been probed through losses in either continuous-time non-Hermitian dynamics of light propagating in optical waveguide array~\cite{Zeunerprl}, or non-unitary discrete-time photonic quantum walks (QWs)~\cite{pxprl}. Interestingly, detected topological invariants in these lossy systems can be associated with underlying non-Hermitian~\cite{Zeunerprl,RLL16} or non-unitary Floquet dynamics~\cite{pxprl}, respectively. These studies reveal topological properties in non-Hermitian or non-unitary settings, and establish a new paradigm of topology that is difficult to access in conventional condensed matter systems~\cite{RLL16,RL09}.

In this work, we report experimental detection of large winding numbers of three and four in photonic non-unitary QWs, which are scalable to feature even higher winding numbers.
By periodical partial measurements on polarization of the photonic walker, we realize multi-step non-unitary QWs in one dimension supporting Floquet topological phases (FTPs). As for two-step non-unitary QWs, partial measurement introduces loss to the quantum-walk dynamics and provides a natural detection channel for FTP winding number~\cite{pxprl,RAA17}.
Whereas FTPs in two-step non-unitary QWs are directly related to those in a lossy Su-Schrieffer-Heeger (SSH) model~\cite{pxprl,RL09}, the multi-step non-unitary QWs here are analogous to adding longer-range hopping terms in the lossy SSH model, which gives rise to higher winding numbers. We directly detect winding numbers of three and four through average displacements, and demonstrate topological phase transitions between FTPs with different topological invariants by probing statistical moments of the walker. We also directly demonstrate, for the first time, the breaking of pseudo-unitary near topological phase boundaries.
Our experimental detection of large winding numbers in non-unitary FTPs offers the exciting prospect of exploring topological phases characterized by large topological invariants in non-unitary or non-Hermitian settings, which will create further opportunities in engineering unconventional topological phenomena using photonics.

\begin{figure*}
\includegraphics[width=\textwidth]{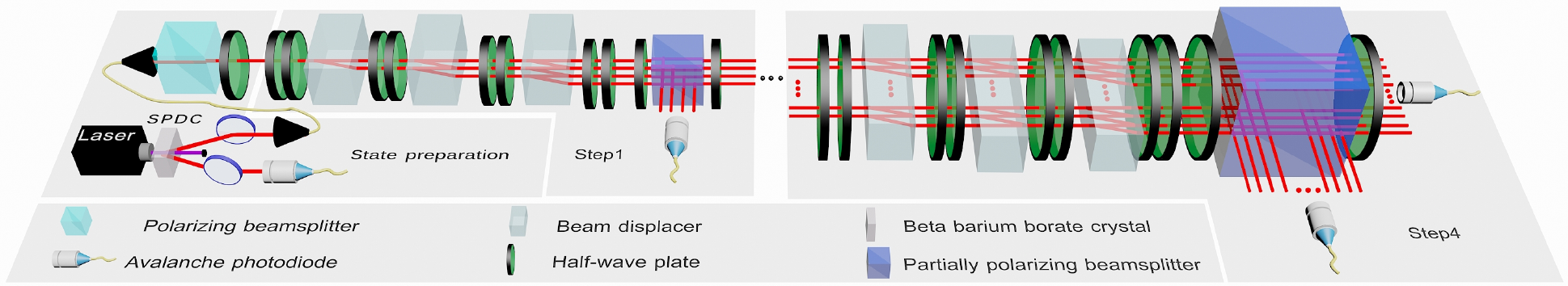}
\caption{We show a three-step non-unitary QW up to $4$ time steps as an example. The photon pair is created via spontaneous parametric downconversion. One photon serves as a trigger. The other photon is projected into the polarization state $\ket{+}$ with a polarizing beamsplitter (PBS) and a half-wave plate (HWP, at $22.5^\circ$) and then proceeds through the quantum-walk interferometric network. The polarization rotation $R$ and the polarization-dependent shift $S$ are realized by two HWPs with certain setting angles depending on the coin parameters $(\theta_1,\theta_2)$ and a beam displacer (BD) whose optical axis is cut so that the photons in $\ket{V}$ are directly transmitted and those in $\ket{H}$ undergo a lateral displacement into a neighboring spatial mode, respectively. A sandwich-type HWP(at $22.5^\circ$)-PPBS-HWP (at $22.5^\circ$) setup (here, PPBS is the abbreviation for a partially polarizing beamsplitter) is used to realize the partial measurement $M_\text{e}$. For horizontally and vertically polarized photons, the transmissivity of the PPBS is $(T_\text{H},T_\text{V})=(1,1-p)$. Finally, the photons are detected by avalanche photodiodes (APDs), in coincidence with the trigger photons. Photon counts give measured probabilities after correcting for relative efficiencies of the different APDs.}
\label{setup}
\end{figure*}

{\it Multi-step non-unitary QWs:---}
We introduce the photonic setup for multi-step non-unitary QWs, where the walker is shifted more than twice at each time step.
We focus on three- and four-step non-unitary QWs in this work.
As illustrated in Fig.~\ref{setup}, the three-step QW is on a one-dimensional homogeneous lattice $L$ ($ L \in \mathbb{Z}$), and the dynamics is governed by the Floquet operator~\cite{CSPRA}
\begin{equation}
\widetilde{U}_3':=MU_3'=MR\left(\frac{\theta_1}{2}\right)SR\left(\theta_2\right)SR\left(\theta_2\right)SR\left(\frac{\theta_1}{2}\right),
\label{eq:U3}
\end{equation}
where the coin operator $R(\theta)$ rotates the coin state by $\theta$ about the $y$-axis, and the conditional position shift operator $S$ moves the walker to the left or right by one step based on the coin state. The coin states are encoded in polarizations of single photons $\{\ket{H},\ket{V}\}$ and the walker states are encoded in their spatial modes. Non-unitary dynamics is enforced by the loss operator
\begin{equation}
M=\one_\text{w}\otimes\left(\ket{+}\bra{+}+\sqrt{1-p}\ket{-}\bra{-}\right),\quad 0<p\leqslant 1,
\end{equation}
where $\ket{\pm}=(\ket{H}\pm\ket{V})/\sqrt{2}$, and $\one_\text{w}=\sum_{L}\ket{x}\bra{x}$ with $x$ denoting the position of the walker. The loss operator is equivalent to performing a partial measurement
$M_e=\one_\text{w}\otimes\sqrt{p}\ket{-}\bra{-}$
in the basis $\{|+\rangle,|-\rangle\}$ at each time step, with $p$ the probability of a successful measurement.

Whereas $R$ and $S$ are implemented by using appropriate wave plates and beam displacers (BDs), the partial measurement operator $M_e$ is realized by a sandwich-type setup involving two half-wave plates (HWPs) and a partially polarizing beamsplitter (PPBS)~\cite{pxprl,supp}. At each measurement step in the quantum-walk dynamics, photons in the state $\ket{-}$ are reflected by the PPBS with probability $p$. Photons are then detected by single-photon avalanche photodiodes (APDs) and lost from the quantum-walk dynamics.

Topological properties in the experimental three-step non-unitary QW are introduced via the effective non-Hermitian Hamiltonian $H'^{(3)}_{\rm eff}$ defined through $\widetilde{U}_3'=\exp\left[-\text{i}H'^{(3)}_{\rm eff}\right]$~\cite{CSPRA,RAA17}.
For the homogeneous single-photon QW considered here, $H'^{(3)}_{\rm eff}(k)=E_k \bm{n}\cdot\bm{\sigma}$ in momentum~$k$ space, with $\bm{\sigma}$ the Pauli vector, $E_k$ the quasienergy spectrum, and $\bm{n}$ the direction of the spinor eigen-vector for each momentum $-\pi<k\leq\pi$. Similar to the case of the two-step non-unitary QW~\cite{pxprl}, the winding number of the three-step QW, which serves as a topological invariant of the system, is the number of times the real component of $\bm{n}$ winds around the $x$-axis as $k$ varies through the first Brillouin zone.

\begin{figure}
\includegraphics[width=0.5\textwidth]{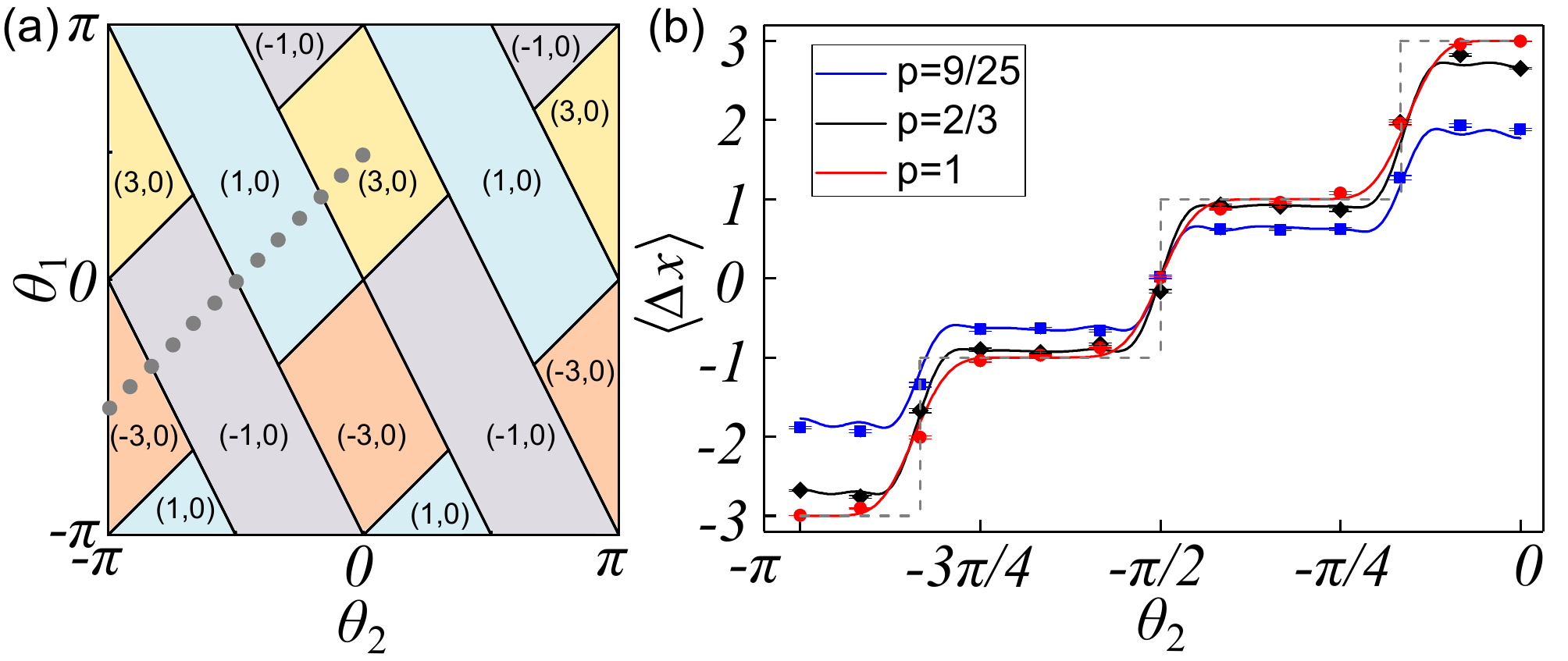}
   \caption{(a) Phase diagram for three-step non-unitary QWs characterized by the topological invariants $(\nu',\nu'')$ as functions of the coin parameters $(\theta_1,\theta_2)$. $(\nu',\nu'')$ are calculated from the Floquet operators $\widetilde{U}_3'$ and $\widetilde{U}_3''$, respectively.
   (b) Measured average displacements of three-step non-unitary QWs corresponding to $\widetilde{U}_3'$ with different loss parameters $p=1, 2/3, 9/25$. Coin parameters vary along the line $\theta_1=\theta_2+\pi/2$, as indicated by dots in Fig.~\ref{displacement3}(a). The dashed curve indicates expected results of infinite-step QWs. The solid curve indicates numerical simulations for QWs with $4$ time steps and the experimental results are presented by dots. Experimental errors are due to photon-counting statistics.}
\label{displacement3}
\end{figure}

For a given FTP with chiral symmetry, two distinct winding numbers $(\nu',\nu'')$ exist for Floquet operators fitted in different time frames~\cite{AO13}. Whereas the corresponding winding number for $\widetilde{U}_3'$ is $\nu'$, $\nu''$ is similarly defined through the winding of the spinor eigen-vector of the non-Hermitian Hamiltonian $H''^{(3)}_{\rm eff}$, where $\widetilde{U}_3''=\exp\left[-\text{i}H''^{(3)}_{\rm eff}\right]$ and
\begin{equation}
\widetilde{U}_3'':=MS_\text{up}R(\theta_2)SR(\theta_1)SR(\theta_2)S_\text{down}.
\label{eq:U3p}
\end{equation}
Here, $S_\text{up}=\sum_x\left(\ket{x+1}\bra{x}\otimes\ket{V}\bra{V}+\ket{x}\bra{x}\otimes\ket{H}\bra{H}\right)$ and $S_\text{down}=\sum_x\left(\ket{x}\bra{x}\otimes\ket{V}\bra{V}+\ket{x-1}\bra{x}\otimes\ket{H}\bra{H}\right)$. Depending on the coin parameters, the absolute value of the winding numbers can take large integer values up to three, as we show in the phase diagram in Fig.~\ref{displacement3}(a).

Similar to three-step QWs, we define four-step non-unitary QWs from constructing the evolution operators
\begin{align}
\widetilde{U}_4'{}^{\left(\right.}{}''{}^{\left)\right.}:=MR\left[\frac{\theta_{1(2)}}{2}\right]SR(0)SR\left[\theta_{2(1)}\right]SR(0)SR\left[\frac{\theta_{1(2)}}{2}\right],
\label{eq:U4}
\end{align}
By analyzing the effective non-Hermitian Hamiltonians $H'^{(4)}_{\rm eff}$ and $H''^{(4)}_{\rm eff}$ respectively associated with the Floquet operators $\widetilde{U}_4'$ and $\widetilde{U}_4''$, it is straightforward to demonstrate that FTPs exist for four-step QWs, which are characterized by integer-valued winding numbers as large as four.
Importantly, both the three- and four-step QWs defined in Eqs.~(\ref{eq:U3}), (\ref{eq:U3p}) and (\ref{eq:U4}) have chiral symmetry in the unitary limit ($p=0$), with the chiral symmetry operator given by $\Gamma=\sigma_x$ as $\Gamma U\Gamma=U^{-1}$~\cite{supp}, where $U$ designates the Floquet operator of the corresponding QW.
Consistent with previous studies~\cite{RLL16}, we find that topological properties of the non-unitary quantum-walk dynamics derive from those in the unitary limit, which are in turn protected by chiral symmetry. Hence, chiral symmetry in the unitary limit is crucial for the perseverance of the FTPs in the non-unitary case ($p>0$). Such a requirement restrains the available forms of non-unitary Floquet operators, of which Eqs.~(\ref{eq:U3}), (\ref{eq:U3p}) and (\ref{eq:U4}) are the most straightforward examples~\cite{supp}.

{\it Detecting topological invariants from losses:---}
In two-step non-unitary QWs, topological invariants can be probed by monitoring losses~\cite{RL09,RAA17,pxprl}. As we experimentally demonstrate and explain, topological invariants of the multi-step non-unitary QWs are determined from losses by measuring average displacement
\begin{equation}
\langle \Delta x\rangle=\sum_x\sum_{t'=1}^{\infty} x P_\text{th}(x,t'),
\end{equation}
for the walker-coin system initialized in the state $\ket{\psi_0}=\ket{x=0}\otimes\ket{+}$. Here, the probability of the walker being detected at $x$ during the $t$-th time step is
\begin{equation}
P_\text{th}(x,t)=\bra{\psi_{t-1}}U_3'^{\dagger}M^\dagger_e \left(\ket{x}\bra{x}\otimes\one_\text{c}\right)M_e U_3'\ket{\psi_{t-1}},
\end{equation}
where $|\psi_t\rangle=(\widetilde{U}'_3)^{t}\ket{\psi_0}$, and $\one_\text{c}$ is a $2\times 2$ identity operator.

To experimentally probe the average displacement in the non-unitary QW with $t$ steps in total, we perform coincidence measurements on the number of the reflected photons $N_\text{R}(x,t')$ ($t'=1,...,t$) at each position successively up to $t$. We then construct the probability \begin{equation}
P_\text{exp}(x,t')=\frac{N_\text{R}(x,t')}{\sum_{x'}\left[\sum^{t}_{t''=1}N_\text{R}(x',t'')+N_\text{T}(x',t)\right]},
\end{equation}
where $N_\text{T}(x,t)$ is the number of transmitted photons at the last step $t$. The average displacement is then
\begin{equation}
\langle \Delta x\rangle_\text{exp}=\sum_x\sum_{t'=1}^{t} x P_\text{exp}(x,t').
\end{equation}

\begin{figure*}
\includegraphics[width=0.85\textwidth]{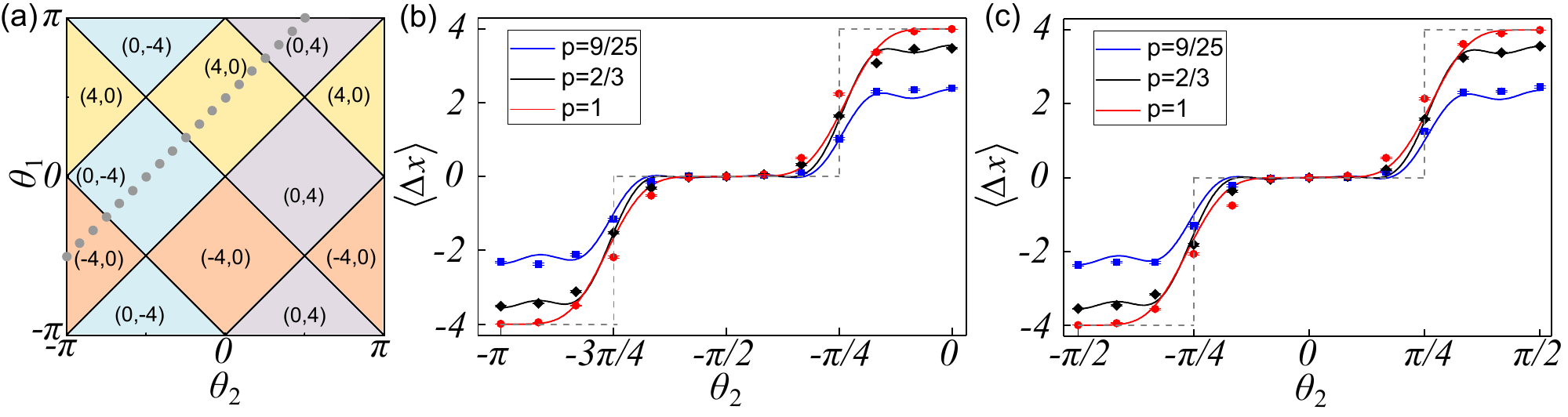}
   \caption{(a) Phase diagram for four-step non-unitary QWs in terms of the topological invariants $(\nu',\nu'')$. $(\nu',\nu'')$ are calculated from the Floquet operators $\widetilde{U}_4'$ and $\widetilde{U}_4''$ respectively. Measured average displacements of four-step non-unitary QWs of $\widetilde{U}_4'$ (b) and $\widetilde{U}_4''$ (c) with different loss parameters $p=1,2/3,9/25$. Coin parameters vary along the line $\theta_1=\theta_2+\pi/2$ as indicated by dots in Fig.~\ref{displacement4}(a). Experimental errors are due to
photon-counting statistics.}
\label{displacement4}
\end{figure*}

To detect topological invariants, we realize three-step non-unitary QWs with three different loss parameters $p=1,2/3,9/25$. The corresponding phase diagram is shown in Fig.~\ref{displacement3}(a), where the topological invariants $(\nu',\nu'')$ are functions of the coin parameters $(\theta_1,\theta_2)$.
Thirteen sets of coin parameters $(\theta_1,\theta_2)$ are chosen along the line $\theta_1=\theta_2+\pi/2$, as indicated in Fig.~\ref{displacement3}(a). The topological invariant $\nu'$ assumes values $-3$, $-1$, $1$ to $3$ along the line, while $\nu''$ is fixed at $0$.
The walker starts from $x = 0$, and the initial coin state is chosen to be $\ket{+}$.

Measured average displacements are shown in Fig.~\ref{displacement3}(b) for the Floquet operator $\widetilde{U}_3'$ (as $\nu''$ is always zero, the average displacements for $\widetilde{U}_3''$ are not shown). These results agree well with the numerical simulations of three-step QWs up to $4$ time steps and demonstrate plateaux close to the quantized values of $\nu'$ calculated for QWs with infinite time steps. We observe that with increasing loss parameter $p$, measured average displacements at a given time step converge faster to the quantized values. This result is consistent with the measurement results for two-step non-unitary QWs~\cite{pxprl} and suggests that the quantum Zeno effect is weak in these systems~\cite{RAA17}. For systems with a strong quantum Zeno effect, $|-\rangle$ becomes effectively unoccupied in the limit of $p=1$, which results in a longer convergence time with increasing $p$.
Meanwhile, regardless of the loss parameter, it takes much longer for the displacements to converge near topological phase transitions, where the topological invariants undergo abrupt changes.

We then implement four-step non-unitary QWs with various loss parameters $p=1,2/3,9/25$. The corresponding phase diagram is shown in Fig.~\ref{displacement4}(a). As the coin parameters vary along the dotted line $\theta_1=\theta_2+\pi/2$ in the phase diagram, the topological invariants $(\nu',\nu'')$ change from $(-4,0)$, $(0,-4)$, $(4,0)$, to $(0,4)$.
The measured average displacements for the operators $\widetilde{U}_4'$ and $\widetilde{U}_4''$ up to $3$ time steps are shown in Figs.~\ref{displacement4}(b) and 3(c), respectively, which agree well with the corresponding numerical simulations.

{\it Confirming the topological phase transitions:---}
We confirm the topological phase boundaries, signaled by jumps of the measured topological invariants, by probing statistical moments~\cite{Cardano2016}.

We define the second statistical moment of the walker after $t$ steps as
\begin{equation}
m_2(t):=\frac{\sum_x x^2 \langle\psi_t|x\rangle\langle x|\otimes \one_\text{c}|\psi_t\rangle}{\sum_x \langle\psi_t|x\rangle\langle x|\otimes \one_\text{c}|\psi_t\rangle}.
\end{equation}
Experimentally, the moment is evaluated from the spatial distribution of the transmitted photons at the last step $t$
\begin{equation}
m^\text{exp}_2(t)=\sum_x x^2 \frac{N_\text{T}(x,t)}{\sum_{x'} N_\text{T}(x',t)}.
\end{equation}

\begin{figure*}
\includegraphics[width=\textwidth]{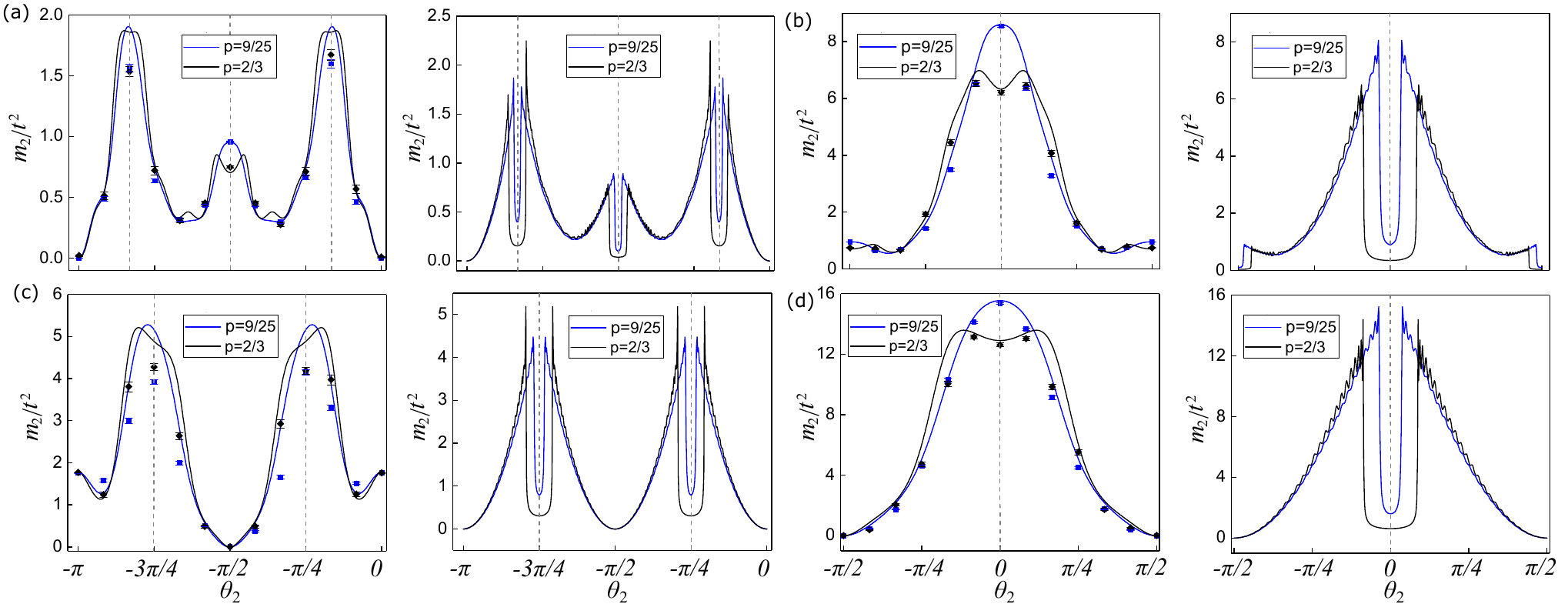}
\caption{Statistical moments $m_2/t^2$ of the walker position distribution for three-step non-unitary QWs governed by $\widetilde{U}'_3$ (upper layer); and for four-step non-unitary QWs governed by $\widetilde{U}'_4$ (lower layer), with the loss parameters $p=9/25,2/3$. (a,c) Coin parameters $(\theta_1,\theta_2)$ are scanned along the dotted lines in the phase diagrams of Figs.~\ref{displacement3}(a) and \ref{displacement4}(a). (b,d) Coin parameters are scanned along $\theta_1=0$. Experimental results of $m_2/t^2$ of up to $4$ time steps ($3$ time steps) and numerical simulations up to $50$ time steps are shown in left and right columns, respectively. The vertical dashed lines indicate locations of topological phase transition from theoretical predictions.
}
\label{moments}
\end{figure*}

In Fig.~\ref{moments}, we plot the measured values for $m^\text{exp}_2(t)/t^2$ of multi-step non-unitary QWs with two different loss parameters $p=2/3,9/25$. In (a,c), coin parameters are scanned along the dotted lines in the phase diagrams. We find reasonable agreement between experimental results and numerical simulations. Here, the measured $m^\text{exp}_2(t)/t^2$ exhibits anomalies near the topological phase transitions. At short time steps, $m^\text{exp}_2(t)/t^2$ peaks at the topological phase boundaries, which is similar to the case in unitary QWs~\cite{Cardano2016,supp}. From numerical simulations, however, we find that in the long-time limit, while the overall peaking structures persist near topological phase boundaries, precipitous dips centered at the boundaries emerge in the second moment.

Further analysis shows that the Floquet operators of both three- and four-step non-unitary QWs can be mapped~\cite{supp}, by a statistical-moment-preserving scaling, to operators with pseudo-unitarity~\cite{M2002a,M2002b,M2004}. Existence of the pseudo-unitarity guarantees the reality of the quasienergy spectra of the effective non-Hermitian Hamiltonian associated with the scaled quantum-walk dynamics. However, in the vicinity of the topological phase boundaries~\cite{supp}, pseudo-unitarity is lost, which gives rise to imaginary-valued quasienergy spectra. Non-unitary QWs in these regions are therefore analogous to those with broken parity-time symmetry, where the long-time spatial distribution of the walker is Gaussian-like rather than ballistic~\cite{PTsymm1,PTsymm2,supp, PTbroken}. This spreading property leads directly to a drop of the second moment in the non-pseudo-unitary regions.

Importantly, the breaking of pseudo-unitarity can be directly observed at small time steps. In Fig.~\ref{moments}(b,d), we show the measured $m^\text{exp}_2(t)/t^2$ for coin parameters scanned along $\theta_1=0$. The aforementioned dips emerge for both the three- and four-step quantum-walks with $p=2/3$. Therefore, under appropriately chosen parameters, as few as four (three) steps are enough to have a clear observation of the breaking of pseudo-unitarity.


%

{\it Final remarks:---}
By detecting winding numbers of three and four, our experiment establishes the feasibility of detecting higher winding numbers through loss in multi-step QW dynamics. We show that as few as four (three) time steps are sufficient to detect winding numbers of three and four under appropriate parameters.
Whereas the implementation and detection of FTPs of even larger winding numbers are possible in our experimental setup by improving the experimental apparatus~\cite{supp}, a promising setup with even better extendability are QWs in the time domain, where by translating the position of the walker into arrival times at the detector, the number of time steps can be significantly increased~\cite{SCP+10,SCP+11,SGP+12}.
Such an extension would significantly enrich the experimentally accessible non-unitary FTPs in one dimension, and would stimulate further studies on dynamic properties of non-unitary FTPs.

Another interesting direction would be the exploration of the relation between FTPs in non-unitary quantum-walk dynamics and those in a parity-time-symmetric configuration~\cite{PTsymm2,PTbroken}. This is particularly relevant due to the existence of hidden pseudo-unitarity in our system, which is intimately connected with the reality of the quasienergy spectrum and hence with parity-time symmetry as well. Our experiment, with its excellent extendibility, opens up the avenue toward a hierarchy of FTPs with large winding numbers, and sheds new light on understanding topological phenomena in non-unitary systems.

\begin{acknowledgments}
This work has been supported by the Natural Science Foundation of China (Grant Nos. 11474049, 11674056, and 11522545) and the Natural Science Foundation of Jiangsu Province (Grant No. BK20160024). LX is supported by the Scientific Research Foundation of Graduate School of Southeast University. WY acknowledges support from the National Key R\&D Program (Grant Nos. 2016YFA0301700, 2017YFA0304800). HO is supported by a Grant-in-Aid for Scientific Research on Innovative Areas ``Topological Materials Science" (Grant Nos JP16H00975 and JP15K21717) and JSPS KAKENHI (Grant Nos JP16K17760 and JP16K05466). BCS acknowledges financial support from the 1000-Talent Plan. LX and XQ contributed equally to this work.\end{acknowledgments}

\bibliography{reference}

\begin{thebibliography}{47}%
\makeatletter
\providecommand \@ifxundefined [1]{%
 \@ifx{#1\undefined}
}%
\providecommand \@ifnum [1]{%
 \ifnum #1\expandafter \@firstoftwo
 \else \expandafter \@secondoftwo
 \fi
}%
\providecommand \@ifx [1]{%
 \ifx #1\expandafter \@firstoftwo
 \else \expandafter \@secondoftwo
 \fi
}%
\providecommand \natexlab [1]{#1}%
\providecommand \enquote  [1]{``#1''}%
\providecommand \bibnamefont  [1]{#1}%
\providecommand \bibfnamefont [1]{#1}%
\providecommand \citenamefont [1]{#1}%
\providecommand \href@noop [0]{\@secondoftwo}%
\providecommand \href [0]{\begingroup \@sanitize@url \@href}%
\providecommand \@href[1]{\@@startlink{#1}\@@href}%
\providecommand \@@href[1]{\endgroup#1\@@endlink}%
\providecommand \@sanitize@url [0]{\catcode `\\12\catcode `\$12\catcode
  `\&12\catcode `\#12\catcode `\^12\catcode `\_12\catcode `\%12\relax}%
\providecommand \@@startlink[1]{}%
\providecommand \@@endlink[0]{}%
\providecommand \url  [0]{\begingroup\@sanitize@url \@url }%
\providecommand \@url [1]{\endgroup\@href {#1}{\urlprefix }}%
\providecommand \urlprefix  [0]{URL }%
\providecommand \Eprint [0]{\href }%
\providecommand \doibase [0]{http://dx.doi.org/}%
\providecommand \selectlanguage [0]{\@gobble}%
\providecommand \bibinfo  [0]{\@secondoftwo}%
\providecommand \bibfield  [0]{\@secondoftwo}%
\providecommand \translation [1]{[#1]}%
\providecommand \BibitemOpen [0]{}%
\providecommand \bibitemStop [0]{}%
\providecommand \bibitemNoStop [0]{.\EOS\space}%
\providecommand \EOS [0]{\spacefactor3000\relax}%
\providecommand \BibitemShut  [1]{\csname bibitem#1\endcsname}%
\let\auto@bib@innerbib\@empty
\bibitem [{\citenamefont {Hasan}\ and\ \citenamefont {Kane}(2010)}]{HKrmp10}%
  \BibitemOpen
  \bibfield  {author} {\bibinfo {author} {\bibfnamefont {M.~Z.}\ \bibnamefont
  {Hasan}}\ and\ \bibinfo {author} {\bibfnamefont {C.~L.}\ \bibnamefont
  {Kane}},\ }\bibfield  {title} {\enquote {\bibinfo {title} {Colloquium:
  topological insulators},}\ }\href@noop {} {\bibfield  {journal} {\bibinfo
  {journal} {Rev. Mod. Phys.}\ }\textbf {\bibinfo {volume} {82}},\ \bibinfo
  {pages} {3045} (\bibinfo {year} {2010})}\BibitemShut {NoStop}%
\bibitem [{\citenamefont {Qi}\ and\ \citenamefont {Zhang}(2011)}]{QZrmp11}%
  \BibitemOpen
  \bibfield  {author} {\bibinfo {author} {\bibfnamefont {X.~L.}\ \bibnamefont
  {Qi}}\ and\ \bibinfo {author} {\bibfnamefont {S.~C.}\ \bibnamefont {Zhang}},\
  }\bibfield  {title} {\enquote {\bibinfo {title} {Topological insulators and
  superconductors},}\ }\href@noop {} {\bibfield  {journal} {\bibinfo  {journal}
  {Rev. Mod. Phys.}\ }\textbf {\bibinfo {volume} {83}},\ \bibinfo {pages}
  {1057} (\bibinfo {year} {2011})}\BibitemShut {NoStop}%
\bibitem [{\citenamefont {Ryu}\ \emph {et~al.}(2010)\citenamefont {Ryu},
  \citenamefont {Schnyder}, \citenamefont {Furusaki},\ and\ \citenamefont
  {Ludwig}}]{Ryu10}%
  \BibitemOpen
  \bibfield  {author} {\bibinfo {author} {\bibfnamefont {S.}~\bibnamefont
  {Ryu}}, \bibinfo {author} {\bibfnamefont {A.~P.}\ \bibnamefont {Schnyder}},
  \bibinfo {author} {\bibfnamefont {A.}~\bibnamefont {Furusaki}}, \ and\
  \bibinfo {author} {\bibfnamefont {A.~W.~W.}\ \bibnamefont {Ludwig}},\
  }\bibfield  {title} {\enquote {\bibinfo {title} {Topological insulators and
  superconductors: tenfold way and dimensional hierarchy},}\ }\href@noop {}
  {\bibfield  {journal} {\bibinfo  {journal} {New J. Phys.}\ }\textbf {\bibinfo
  {volume} {12}},\ \bibinfo {pages} {065010} (\bibinfo {year}
  {2010})}\BibitemShut {NoStop}%
\bibitem [{\citenamefont {Teo}\ and\ \citenamefont {Kane}(2010)}]{Kane10}%
  \BibitemOpen
  \bibfield  {author} {\bibinfo {author} {\bibfnamefont {J.~C.~Y.}\
  \bibnamefont {Teo}}\ and\ \bibinfo {author} {\bibfnamefont {C.~L.}\
  \bibnamefont {Kane}},\ }\bibfield  {title} {\enquote {\bibinfo {title}
  {Topological defects and gapless modes in insulators and superconductors},}\
  }\href@noop {} {\bibfield  {journal} {\bibinfo  {journal} {Phys. Rev. B}\
  }\textbf {\bibinfo {volume} {82}},\ \bibinfo {pages} {115120} (\bibinfo
  {year} {2010})}\BibitemShut {NoStop}%
\bibitem [{\citenamefont {Chen}\ \emph {et~al.}(2009)\citenamefont {Chen},
  \citenamefont {Analytis}, \citenamefont {Chu}, \citenamefont {Liu},
  \citenamefont {Mo}, \citenamefont {Qi}, \citenamefont {Zhang}, \citenamefont
  {Lu}, \citenamefont {Dai}, \citenamefont {Fang}, \citenamefont {Zhang},
  \citenamefont {Fisher}, \citenamefont {Hussain},\ and\ \citenamefont
  {Shen}}]{CYL09}%
  \BibitemOpen
  \bibfield  {author} {\bibinfo {author} {\bibfnamefont {Y.~L.}\ \bibnamefont
  {Chen}}, \bibinfo {author} {\bibfnamefont {J.~G.}\ \bibnamefont {Analytis}},
  \bibinfo {author} {\bibfnamefont {J.~H.}\ \bibnamefont {Chu}}, \bibinfo
  {author} {\bibfnamefont {Z.~K.}\ \bibnamefont {Liu}}, \bibinfo {author}
  {\bibfnamefont {S.~K.}\ \bibnamefont {Mo}}, \bibinfo {author} {\bibfnamefont
  {X.~L.}\ \bibnamefont {Qi}}, \bibinfo {author} {\bibfnamefont {H.~J.}\
  \bibnamefont {Zhang}}, \bibinfo {author} {\bibfnamefont {D.~H.}\ \bibnamefont
  {Lu}}, \bibinfo {author} {\bibfnamefont {X.}~\bibnamefont {Dai}}, \bibinfo
  {author} {\bibfnamefont {Z.}~\bibnamefont {Fang}}, \bibinfo {author}
  {\bibfnamefont {S.~C.}\ \bibnamefont {Zhang}}, \bibinfo {author}
  {\bibfnamefont {I.~R.}\ \bibnamefont {Fisher}}, \bibinfo {author}
  {\bibfnamefont {Z.}~\bibnamefont {Hussain}}, \ and\ \bibinfo {author}
  {\bibfnamefont {Z.~X.}\ \bibnamefont {Shen}},\ }\bibfield  {title} {\enquote
  {\bibinfo {title} {Experimental realization of a three-dimensional
  topological insulator, {B}i2{T}e3},}\ }\href@noop {} {\bibfield  {journal}
  {\bibinfo  {journal} {Science}\ }\textbf {\bibinfo {volume} {325}},\ \bibinfo
  {pages} {178--181} (\bibinfo {year} {2009})}\BibitemShut {NoStop}%
\bibitem [{\citenamefont {Xia}\ \emph {et~al.}(2009)\citenamefont {Xia},
  \citenamefont {Qian}, \citenamefont {Hsieh}, \citenamefont {Wray},
  \citenamefont {Pal}, \citenamefont {Lin}, \citenamefont {Bansil},
  \citenamefont {Grauer}, \citenamefont {Hor}, \citenamefont {Cava},\ and\
  \citenamefont {Hasan}}]{Xia09}%
  \BibitemOpen
  \bibfield  {author} {\bibinfo {author} {\bibfnamefont {Y.}~\bibnamefont
  {Xia}}, \bibinfo {author} {\bibfnamefont {D.}~\bibnamefont {Qian}}, \bibinfo
  {author} {\bibfnamefont {D.}~\bibnamefont {Hsieh}}, \bibinfo {author}
  {\bibfnamefont {L.}~\bibnamefont {Wray}}, \bibinfo {author} {\bibfnamefont
  {A.}~\bibnamefont {Pal}}, \bibinfo {author} {\bibfnamefont {H.}~\bibnamefont
  {Lin}}, \bibinfo {author} {\bibfnamefont {A.}~\bibnamefont {Bansil}},
  \bibinfo {author} {\bibfnamefont {D.~H.}\ \bibnamefont {Grauer}}, \bibinfo
  {author} {\bibfnamefont {Y.~S.}\ \bibnamefont {Hor}}, \bibinfo {author}
  {\bibfnamefont {R.~J.}\ \bibnamefont {Cava}}, \ and\ \bibinfo {author}
  {\bibfnamefont {M.~Z.}\ \bibnamefont {Hasan}},\ }\bibfield  {title} {\enquote
  {\bibinfo {title} {Observation of a large-gap topological-insulator class
  with a single dirac cone on the surface},}\ }\href@noop {} {\bibfield
  {journal} {\bibinfo  {journal} {Nat. Phys.}\ }\textbf {\bibinfo {volume}
  {5}},\ \bibinfo {pages} {398} (\bibinfo {year} {2009})}\BibitemShut {NoStop}%
\bibitem [{\citenamefont {Wang}\ \emph {et~al.}(2009)\citenamefont {Wang},
  \citenamefont {Chong}, \citenamefont {Joannopoulos},\ and\ \citenamefont
  {Solja{\v{c}}i{\'c}}}]{WCJS09}%
  \BibitemOpen
  \bibfield  {author} {\bibinfo {author} {\bibfnamefont {Z.}~\bibnamefont
  {Wang}}, \bibinfo {author} {\bibfnamefont {Y.}~\bibnamefont {Chong}},
  \bibinfo {author} {\bibfnamefont {J.~D.}\ \bibnamefont {Joannopoulos}}, \
  and\ \bibinfo {author} {\bibfnamefont {M.}~\bibnamefont
  {Solja{\v{c}}i{\'c}}},\ }\bibfield  {title} {\enquote {\bibinfo {title}
  {Observation of unidirectional backscattering-immune topological
  electromagnetic states},}\ }\href@noop {} {\bibfield  {journal} {\bibinfo
  {journal} {Nature}\ }\textbf {\bibinfo {volume} {461}},\ \bibinfo {pages}
  {772} (\bibinfo {year} {2009})}\BibitemShut {NoStop}%
\bibitem [{\citenamefont {Lu}\ \emph {et~al.}(2014)\citenamefont {Lu},
  \citenamefont {Joannopoulos},\ and\ \citenamefont
  {Solja{\v{c}}i{\'c}}}]{LJS14}%
  \BibitemOpen
  \bibfield  {author} {\bibinfo {author} {\bibfnamefont {L.}~\bibnamefont
  {Lu}}, \bibinfo {author} {\bibfnamefont {J.~D.}\ \bibnamefont
  {Joannopoulos}}, \ and\ \bibinfo {author} {\bibfnamefont {M.}~\bibnamefont
  {Solja{\v{c}}i{\'c}}},\ }\bibfield  {title} {\enquote {\bibinfo {title}
  {Topological photonics},}\ }\href@noop {} {\bibfield  {journal} {\bibinfo
  {journal} {Nat. Photon.}\ }\textbf {\bibinfo {volume} {8}},\ \bibinfo {pages}
  {821} (\bibinfo {year} {2014})}\BibitemShut {NoStop}%
\bibitem [{\citenamefont {Kitagawa}\ \emph {et~al.}(2012)\citenamefont
  {Kitagawa}, \citenamefont {Broome}, \citenamefont {Fedrizzi}, \citenamefont
  {Rudner}, \citenamefont {Berg}, \citenamefont {Kassal}, \citenamefont
  {Aspuru-Guzik}, \citenamefont {Demler},\ and\ \citenamefont {White}}]{KB+12}%
  \BibitemOpen
  \bibfield  {author} {\bibinfo {author} {\bibfnamefont {T.}~\bibnamefont
  {Kitagawa}}, \bibinfo {author} {\bibfnamefont {M.~A.}\ \bibnamefont
  {Broome}}, \bibinfo {author} {\bibfnamefont {A.}~\bibnamefont {Fedrizzi}},
  \bibinfo {author} {\bibfnamefont {M.~S.}\ \bibnamefont {Rudner}}, \bibinfo
  {author} {\bibfnamefont {E.}~\bibnamefont {Berg}}, \bibinfo {author}
  {\bibfnamefont {I.}~\bibnamefont {Kassal}}, \bibinfo {author} {\bibfnamefont
  {A.}~\bibnamefont {Aspuru-Guzik}}, \bibinfo {author} {\bibfnamefont
  {E.}~\bibnamefont {Demler}}, \ and\ \bibinfo {author} {\bibfnamefont {A.~G.}\
  \bibnamefont {White}},\ }\bibfield  {title} {\enquote {\bibinfo {title}
  {Observation of topologically protected bound states in photonic quantum
  walks},}\ }\href@noop {} {\bibfield  {journal} {\bibinfo  {journal} {Nat.
  Commun.}\ }\textbf {\bibinfo {volume} {3}},\ \bibinfo {pages} {882} (\bibinfo
  {year} {2012})}\BibitemShut {NoStop}%
\bibitem [{\citenamefont {Skirlo}\ \emph {et~al.}(2015)\citenamefont {Skirlo},
  \citenamefont {Lu}, \citenamefont {Igarashi}, \citenamefont {Yan},
  \citenamefont {Joannopoulos},\ and\ \citenamefont
  {Solja{\v{c}}i{\'c}}}]{LL15}%
  \BibitemOpen
  \bibfield  {author} {\bibinfo {author} {\bibfnamefont {S.~A.}\ \bibnamefont
  {Skirlo}}, \bibinfo {author} {\bibfnamefont {L.}~\bibnamefont {Lu}}, \bibinfo
  {author} {\bibfnamefont {Y.}~\bibnamefont {Igarashi}}, \bibinfo {author}
  {\bibfnamefont {Q.}~\bibnamefont {Yan}}, \bibinfo {author} {\bibfnamefont
  {J.}~\bibnamefont {Joannopoulos}}, \ and\ \bibinfo {author} {\bibfnamefont
  {M.}~\bibnamefont {Solja{\v{c}}i{\'c}}},\ }\bibfield  {title} {\enquote
  {\bibinfo {title} {Experimental observation of large {C}hern numbers in
  photonic crystals},}\ }\href@noop {} {\bibfield  {journal} {\bibinfo
  {journal} {Phys. Rev. Lett.}\ }\textbf {\bibinfo {volume} {115}},\ \bibinfo
  {pages} {253901} (\bibinfo {year} {2015})}\BibitemShut {NoStop}%
\bibitem [{\citenamefont {Cardano}\ \emph {et~al.}(2016)\citenamefont
  {Cardano}, \citenamefont {Maffei}, \citenamefont {Massa}, \citenamefont
  {Piccirillo}, \citenamefont {De~Lisio}, \citenamefont {De~Filippis},
  \citenamefont {Cataudella}, \citenamefont {Santamato},\ and\ \citenamefont
  {Marrucci}}]{Cardano2016}%
  \BibitemOpen
  \bibfield  {author} {\bibinfo {author} {\bibfnamefont {F.}~\bibnamefont
  {Cardano}}, \bibinfo {author} {\bibfnamefont {M.}~\bibnamefont {Maffei}},
  \bibinfo {author} {\bibfnamefont {F.}~\bibnamefont {Massa}}, \bibinfo
  {author} {\bibfnamefont {B.}~\bibnamefont {Piccirillo}}, \bibinfo {author}
  {\bibfnamefont {C.}~\bibnamefont {De~Lisio}}, \bibinfo {author}
  {\bibfnamefont {G.}~\bibnamefont {De~Filippis}}, \bibinfo {author}
  {\bibfnamefont {V.}~\bibnamefont {Cataudella}}, \bibinfo {author}
  {\bibfnamefont {E.}~\bibnamefont {Santamato}}, \ and\ \bibinfo {author}
  {\bibfnamefont {L.}~\bibnamefont {Marrucci}},\ }\bibfield  {title} {\enquote
  {\bibinfo {title} {Statistical moments of quantum-walk dynamics reveal
  topological quantum transitions},}\ }\href@noop {} {\bibfield  {journal}
  {\bibinfo  {journal} {Nat. Commun.}\ }\textbf {\bibinfo {volume} {7}},\
  \bibinfo {pages} {11439} (\bibinfo {year} {2016})}\BibitemShut {NoStop}%
\bibitem [{\citenamefont {Xiao}\ \emph {et~al.}(2017)\citenamefont {Xiao},
  \citenamefont {Zhan}, \citenamefont {Bian}, \citenamefont {Wang},
  \citenamefont {Zhang}, \citenamefont {Wang}, \citenamefont {Li},
  \citenamefont {Mochizuki}, \citenamefont {Kim}, \citenamefont {Kawakami},
  \citenamefont {Yi}, \citenamefont {Obuse}, \citenamefont {Sanders},\ and\
  \citenamefont {Xue}}]{PTsymm2}%
  \BibitemOpen
  \bibfield  {author} {\bibinfo {author} {\bibfnamefont {L.}~\bibnamefont
  {Xiao}}, \bibinfo {author} {\bibfnamefont {X.}~\bibnamefont {Zhan}}, \bibinfo
  {author} {\bibfnamefont {Z.~H.}\ \bibnamefont {Bian}}, \bibinfo {author}
  {\bibfnamefont {K.~K.}\ \bibnamefont {Wang}}, \bibinfo {author}
  {\bibfnamefont {X.}~\bibnamefont {Zhang}}, \bibinfo {author} {\bibfnamefont
  {X.~P.}\ \bibnamefont {Wang}}, \bibinfo {author} {\bibfnamefont
  {J.}~\bibnamefont {Li}}, \bibinfo {author} {\bibfnamefont {K.}~\bibnamefont
  {Mochizuki}}, \bibinfo {author} {\bibfnamefont {D.}~\bibnamefont {Kim}},
  \bibinfo {author} {\bibfnamefont {N.}~\bibnamefont {Kawakami}}, \bibinfo
  {author} {\bibfnamefont {W.}~\bibnamefont {Yi}}, \bibinfo {author}
  {\bibfnamefont {H.}~\bibnamefont {Obuse}}, \bibinfo {author} {\bibfnamefont
  {B.~C.}\ \bibnamefont {Sanders}}, \ and\ \bibinfo {author} {\bibfnamefont
  {P.}~\bibnamefont {Xue}},\ }\bibfield  {title} {\enquote {\bibinfo {title}
  {Observation of topological edge states in parity-time-symmetric quantum
  walks},}\ }\href@noop {} {\bibfield  {journal} {\bibinfo  {journal} {Nat.
  Phys.}\ }\textbf {\bibinfo {volume} {13}},\ \bibinfo {pages} {1117} (\bibinfo
  {year} {2017})}\BibitemShut {NoStop}%
\bibitem [{\citenamefont {Poli}\ \emph {et~al.}(2015)\citenamefont {Poli},
  \citenamefont {Bellec}, \citenamefont {Kuhl}, \citenamefont {Mortessagne},\
  and\ \citenamefont {Schomerus}}]{PBKMS15}%
  \BibitemOpen
  \bibfield  {author} {\bibinfo {author} {\bibfnamefont {C.}~\bibnamefont
  {Poli}}, \bibinfo {author} {\bibfnamefont {M.}~\bibnamefont {Bellec}},
  \bibinfo {author} {\bibfnamefont {U.}~\bibnamefont {Kuhl}}, \bibinfo {author}
  {\bibfnamefont {F.}~\bibnamefont {Mortessagne}}, \ and\ \bibinfo {author}
  {\bibfnamefont {H.}~\bibnamefont {Schomerus}},\ }\bibfield  {title} {\enquote
  {\bibinfo {title} {Selective enhancement of topologically induced interface
  states in a dielectric resonator chain},}\ }\href@noop {} {\bibfield
  {journal} {\bibinfo  {journal} {Nat. Commun.}\ }\textbf {\bibinfo {volume}
  {6}},\ \bibinfo {pages} {6710} (\bibinfo {year} {2015})}\BibitemShut
  {NoStop}%
\bibitem [{\citenamefont {Bellec}\ \emph {et~al.}(2013)\citenamefont {Bellec},
  \citenamefont {Kuhl}, \citenamefont {Montambaux},\ and\ \citenamefont
  {Mortessagne}}]{BKMM13}%
  \BibitemOpen
  \bibfield  {author} {\bibinfo {author} {\bibfnamefont {M.}~\bibnamefont
  {Bellec}}, \bibinfo {author} {\bibfnamefont {U.}~\bibnamefont {Kuhl}},
  \bibinfo {author} {\bibfnamefont {G.}~\bibnamefont {Montambaux}}, \ and\
  \bibinfo {author} {\bibfnamefont {F.}~\bibnamefont {Mortessagne}},\
  }\bibfield  {title} {\enquote {\bibinfo {title} {Topological transition of
  {D}irac points in a microwave experiment},}\ }\href@noop {} {\bibfield
  {journal} {\bibinfo  {journal} {Phys. Rev. Lett.}\ }\textbf {\bibinfo
  {volume} {110}},\ \bibinfo {pages} {033902} (\bibinfo {year}
  {2013})}\BibitemShut {NoStop}%
\bibitem [{\citenamefont {Hu}\ \emph {et~al.}(2015)\citenamefont {Hu},
  \citenamefont {Pillay}, \citenamefont {Wu}, \citenamefont {Pasek},
  \citenamefont {Shum},\ and\ \citenamefont {Chong}}]{Chong15}%
  \BibitemOpen
  \bibfield  {author} {\bibinfo {author} {\bibfnamefont {W.}~\bibnamefont
  {Hu}}, \bibinfo {author} {\bibfnamefont {J.~C.}\ \bibnamefont {Pillay}},
  \bibinfo {author} {\bibfnamefont {K.}~\bibnamefont {Wu}}, \bibinfo {author}
  {\bibfnamefont {M.}~\bibnamefont {Pasek}}, \bibinfo {author} {\bibfnamefont
  {P.~P.}\ \bibnamefont {Shum}}, \ and\ \bibinfo {author} {\bibfnamefont
  {Y.~D.}\ \bibnamefont {Chong}},\ }\bibfield  {title} {\enquote {\bibinfo
  {title} {Measurement of a topological edge invariant in a microwave
  network},}\ }\href@noop {} {\bibfield  {journal} {\bibinfo  {journal} {Phys.
  Rev. X}\ }\textbf {\bibinfo {volume} {5}},\ \bibinfo {pages} {011012}
  (\bibinfo {year} {2015})}\BibitemShut {NoStop}%
\bibitem [{\citenamefont {S{\"u}sstrunk}\ and\ \citenamefont
  {Huber}(2015)}]{Hubers}%
  \BibitemOpen
  \bibfield  {author} {\bibinfo {author} {\bibfnamefont {R.}~\bibnamefont
  {S{\"u}sstrunk}}\ and\ \bibinfo {author} {\bibfnamefont {S.~D.}\ \bibnamefont
  {Huber}},\ }\bibfield  {title} {\enquote {\bibinfo {title} {Observation of
  phononic helical edge states in a mechanical topological insulator},}\
  }\href@noop {} {\bibfield  {journal} {\bibinfo  {journal} {Science}\ }\textbf
  {\bibinfo {volume} {349}},\ \bibinfo {pages} {47--50} (\bibinfo {year}
  {2015})}\BibitemShut {NoStop}%
\bibitem [{\citenamefont {Fleury}\ \emph {et~al.}(2016)\citenamefont {Fleury},
  \citenamefont {Khanikaev},\ and\ \citenamefont {Alu}}]{Khanikaevnc}%
  \BibitemOpen
  \bibfield  {author} {\bibinfo {author} {\bibfnamefont {R.}~\bibnamefont
  {Fleury}}, \bibinfo {author} {\bibfnamefont {A.~B.}\ \bibnamefont
  {Khanikaev}}, \ and\ \bibinfo {author} {\bibfnamefont {A.}~\bibnamefont
  {Alu}},\ }\bibfield  {title} {\enquote {\bibinfo {title} {Floquet topological
  insulators for sound},}\ }\href@noop {} {\bibfield  {journal} {\bibinfo
  {journal} {Nat. Commun.}\ }\textbf {\bibinfo {volume} {7}},\ \bibinfo {pages}
  {11744} (\bibinfo {year} {2016})}\BibitemShut {NoStop}%
\bibitem [{\citenamefont {Jotzu}\ \emph {et~al.}(2014)\citenamefont {Jotzu},
  \citenamefont {Messer}, \citenamefont {Desbuquois}, \citenamefont {Lebrat},
  \citenamefont {Uehlinger}, \citenamefont {Greif},\ and\ \citenamefont
  {Esslinger}}]{ETHcoldatom14}%
  \BibitemOpen
  \bibfield  {author} {\bibinfo {author} {\bibfnamefont {G.}~\bibnamefont
  {Jotzu}}, \bibinfo {author} {\bibfnamefont {M.}~\bibnamefont {Messer}},
  \bibinfo {author} {\bibfnamefont {R.}~\bibnamefont {Desbuquois}}, \bibinfo
  {author} {\bibfnamefont {M.}~\bibnamefont {Lebrat}}, \bibinfo {author}
  {\bibfnamefont {T.}~\bibnamefont {Uehlinger}}, \bibinfo {author}
  {\bibfnamefont {D.}~\bibnamefont {Greif}}, \ and\ \bibinfo {author}
  {\bibfnamefont {T.}~\bibnamefont {Esslinger}},\ }\bibfield  {title} {\enquote
  {\bibinfo {title} {Experimental realization of the topological {H}aldane
  model with ultracold fermions},}\ }\href@noop {} {\bibfield  {journal}
  {\bibinfo  {journal} {Nature}\ }\textbf {\bibinfo {volume} {515}},\ \bibinfo
  {pages} {237} (\bibinfo {year} {2014})}\BibitemShut {NoStop}%
\bibitem [{\citenamefont {Leder}\ \emph {et~al.}(2016)\citenamefont {Leder},
  \citenamefont {Grossert}, \citenamefont {Sitta}, \citenamefont {Genske},
  \citenamefont {Rosch},\ and\ \citenamefont {Weitz}}]{Weitz16}%
  \BibitemOpen
  \bibfield  {author} {\bibinfo {author} {\bibfnamefont {M.}~\bibnamefont
  {Leder}}, \bibinfo {author} {\bibfnamefont {C.}~\bibnamefont {Grossert}},
  \bibinfo {author} {\bibfnamefont {L.}~\bibnamefont {Sitta}}, \bibinfo
  {author} {\bibfnamefont {M.}~\bibnamefont {Genske}}, \bibinfo {author}
  {\bibfnamefont {A.}~\bibnamefont {Rosch}}, \ and\ \bibinfo {author}
  {\bibfnamefont {M.}~\bibnamefont {Weitz}},\ }\bibfield  {title} {\enquote
  {\bibinfo {title} {Real-space imaging of a topologically protected edge state
  with ultracold atoms in an amplitude-chirped optical lattice},}\ }\href@noop
  {} {\bibfield  {journal} {\bibinfo  {journal} {Nat. Commun.}\ }\textbf
  {\bibinfo {volume} {7}},\ \bibinfo {pages} {13112} (\bibinfo {year}
  {2016})}\BibitemShut {NoStop}%
\bibitem [{\citenamefont {Meier}\ \emph {et~al.}(2016)\citenamefont {Meier},
  \citenamefont {An},\ and\ \citenamefont {Gadway}}]{Gadway16}%
  \BibitemOpen
  \bibfield  {author} {\bibinfo {author} {\bibfnamefont {E.~J.}\ \bibnamefont
  {Meier}}, \bibinfo {author} {\bibfnamefont {F.~A.}\ \bibnamefont {An}}, \
  and\ \bibinfo {author} {\bibfnamefont {B.}~\bibnamefont {Gadway}},\
  }\bibfield  {title} {\enquote {\bibinfo {title} {Observation of the
  topological soliton state in the {S}u-{S}chrieffer-{H}eeger model},}\
  }\href@noop {} {\bibfield  {journal} {\bibinfo  {journal} {Nat. Commun.}\
  }\textbf {\bibinfo {volume} {7}},\ \bibinfo {pages} {13986} (\bibinfo {year}
  {2016})}\BibitemShut {NoStop}%
\bibitem [{\citenamefont {Atala}\ \emph {et~al.}(2013)\citenamefont {Atala},
  \citenamefont {Aidelsburger}, \citenamefont {Barreiro}, \citenamefont
  {Abanin}, \citenamefont {Kitagawa}, \citenamefont {Demler},\ and\
  \citenamefont {Bloch}}]{Bloch13}%
  \BibitemOpen
  \bibfield  {author} {\bibinfo {author} {\bibfnamefont {M.}~\bibnamefont
  {Atala}}, \bibinfo {author} {\bibfnamefont {M.}~\bibnamefont {Aidelsburger}},
  \bibinfo {author} {\bibfnamefont {J.~T.}\ \bibnamefont {Barreiro}}, \bibinfo
  {author} {\bibfnamefont {D.}~\bibnamefont {Abanin}}, \bibinfo {author}
  {\bibfnamefont {T.}~\bibnamefont {Kitagawa}}, \bibinfo {author}
  {\bibfnamefont {E.}~\bibnamefont {Demler}}, \ and\ \bibinfo {author}
  {\bibfnamefont {I.}~\bibnamefont {Bloch}},\ }\bibfield  {title} {\enquote
  {\bibinfo {title} {Direct measurement of the {Z}ak phase in topological
  {B}loch bands},}\ }\href@noop {} {\bibfield  {journal} {\bibinfo  {journal}
  {Nat. Phys.}\ }\textbf {\bibinfo {volume} {9}},\ \bibinfo {pages} {795}
  (\bibinfo {year} {2013})}\BibitemShut {NoStop}%
\bibitem [{\citenamefont {Aidelsburger}\ \emph {et~al.}(2015)\citenamefont
  {Aidelsburger}, \citenamefont {Lohse}, \citenamefont {Schweizer},
  \citenamefont {Atala}, \citenamefont {Barreiro}, \citenamefont {Nascimbene},
  \citenamefont {Cooper}, \citenamefont {Bloch},\ and\ \citenamefont
  {Goldman}}]{CBG15}%
  \BibitemOpen
  \bibfield  {author} {\bibinfo {author} {\bibfnamefont {M.}~\bibnamefont
  {Aidelsburger}}, \bibinfo {author} {\bibfnamefont {M.}~\bibnamefont {Lohse}},
  \bibinfo {author} {\bibfnamefont {C.}~\bibnamefont {Schweizer}}, \bibinfo
  {author} {\bibfnamefont {M.}~\bibnamefont {Atala}}, \bibinfo {author}
  {\bibfnamefont {J.~T.}\ \bibnamefont {Barreiro}}, \bibinfo {author}
  {\bibfnamefont {S.}~\bibnamefont {Nascimbene}}, \bibinfo {author}
  {\bibfnamefont {N.~R.}\ \bibnamefont {Cooper}}, \bibinfo {author}
  {\bibfnamefont {I.}~\bibnamefont {Bloch}}, \ and\ \bibinfo {author}
  {\bibfnamefont {N.}~\bibnamefont {Goldman}},\ }\bibfield  {title} {\enquote
  {\bibinfo {title} {Measuring the {C}hern number of {H}ofstadter bands with
  ultracold bosonic atoms},}\ }\href@noop {} {\bibfield  {journal} {\bibinfo
  {journal} {Nat. Phys.}\ }\textbf {\bibinfo {volume} {11}},\ \bibinfo {pages}
  {162} (\bibinfo {year} {2015})}\BibitemShut {NoStop}%
\bibitem [{\citenamefont {Fl{\"a}schner}\ \emph {et~al.}(2016)\citenamefont
  {Fl{\"a}schner}, \citenamefont {Rem}, \citenamefont {Tarnowski},
  \citenamefont {Vogel}, \citenamefont {L{\"u}hmann}, \citenamefont
  {Sengstock},\ and\ \citenamefont {Weitenberg}}]{Weitenberg2016}%
  \BibitemOpen
  \bibfield  {author} {\bibinfo {author} {\bibfnamefont {N.}~\bibnamefont
  {Fl{\"a}schner}}, \bibinfo {author} {\bibfnamefont {B.~S.}\ \bibnamefont
  {Rem}}, \bibinfo {author} {\bibfnamefont {M.}~\bibnamefont {Tarnowski}},
  \bibinfo {author} {\bibfnamefont {D.}~\bibnamefont {Vogel}}, \bibinfo
  {author} {\bibfnamefont {D.~S.}\ \bibnamefont {L{\"u}hmann}}, \bibinfo
  {author} {\bibfnamefont {K.}~\bibnamefont {Sengstock}}, \ and\ \bibinfo
  {author} {\bibfnamefont {C.}~\bibnamefont {Weitenberg}},\ }\bibfield  {title}
  {\enquote {\bibinfo {title} {Experimental reconstruction of the {B}erry
  curvature in a {F}loquet {B}loch band},}\ }\href@noop {} {\bibfield
  {journal} {\bibinfo  {journal} {Science}\ }\textbf {\bibinfo {volume}
  {352}},\ \bibinfo {pages} {1091--1094} (\bibinfo {year} {2016})}\BibitemShut
  {NoStop}%
\bibitem [{\citenamefont {Ramasesh}\ \emph {et~al.}(2017)\citenamefont
  {Ramasesh}, \citenamefont {Flurin}, \citenamefont {Rudner}, \citenamefont
  {Siddiqi},\ and\ \citenamefont {Yao}}]{RFR+17}%
  \BibitemOpen
  \bibfield  {author} {\bibinfo {author} {\bibfnamefont {V.~V.}\ \bibnamefont
  {Ramasesh}}, \bibinfo {author} {\bibfnamefont {E.}~\bibnamefont {Flurin}},
  \bibinfo {author} {\bibfnamefont {M.}~\bibnamefont {Rudner}}, \bibinfo
  {author} {\bibfnamefont {I.}~\bibnamefont {Siddiqi}}, \ and\ \bibinfo
  {author} {\bibfnamefont {N.~Y.}\ \bibnamefont {Yao}},\ }\bibfield  {title}
  {\enquote {\bibinfo {title} {Direct probe of topological invariants using
  bloch oscillating quantum walks},}\ }\href@noop {} {\bibfield  {journal}
  {\bibinfo  {journal} {Phys. Rev. Lett.}\ }\textbf {\bibinfo {volume} {118}},\
  \bibinfo {pages} {130501} (\bibinfo {year} {2017})}\BibitemShut {NoStop}%
\bibitem [{\citenamefont {Flurin}\ \emph {et~al.}(2017)\citenamefont {Flurin},
  \citenamefont {Ramasesh}, \citenamefont {Hacohen-Gourgy}, \citenamefont
  {Martin}, \citenamefont {Yao},\ and\ \citenamefont {Siddiqi}}]{FRH+16}%
  \BibitemOpen
  \bibfield  {author} {\bibinfo {author} {\bibfnamefont {E.}~\bibnamefont
  {Flurin}}, \bibinfo {author} {\bibfnamefont {V.~V.}\ \bibnamefont
  {Ramasesh}}, \bibinfo {author} {\bibfnamefont {S.}~\bibnamefont
  {Hacohen-Gourgy}}, \bibinfo {author} {\bibfnamefont {L.~S.}\ \bibnamefont
  {Martin}}, \bibinfo {author} {\bibfnamefont {N.~Y.}\ \bibnamefont {Yao}}, \
  and\ \bibinfo {author} {\bibfnamefont {I.}~\bibnamefont {Siddiqi}},\
  }\bibfield  {title} {\enquote {\bibinfo {title} {Observing topological
  invariants using quantum walks in superconducting circuits},}\ }\href@noop {}
  {\bibfield  {journal} {\bibinfo  {journal} {Phys. Rev. X}\ }\textbf {\bibinfo
  {volume} {7}},\ \bibinfo {pages} {031023} (\bibinfo {year}
  {2017})}\BibitemShut {NoStop}%
\bibitem [{\citenamefont {Zhan}\ \emph {et~al.}(2017)\citenamefont {Zhan},
  \citenamefont {Xiao}, \citenamefont {Bian}, \citenamefont {Wang},
  \citenamefont {Qiu}, \citenamefont {Sanders}, \citenamefont {Yi},\ and\
  \citenamefont {Xue}}]{pxprl}%
  \BibitemOpen
  \bibfield  {author} {\bibinfo {author} {\bibfnamefont {X.}~\bibnamefont
  {Zhan}}, \bibinfo {author} {\bibfnamefont {L.}~\bibnamefont {Xiao}}, \bibinfo
  {author} {\bibfnamefont {Z.~H.}\ \bibnamefont {Bian}}, \bibinfo {author}
  {\bibfnamefont {K.~K.}\ \bibnamefont {Wang}}, \bibinfo {author}
  {\bibfnamefont {X.~Z.}\ \bibnamefont {Qiu}}, \bibinfo {author} {\bibfnamefont
  {B.~C.}\ \bibnamefont {Sanders}}, \bibinfo {author} {\bibfnamefont
  {W.}~\bibnamefont {Yi}}, \ and\ \bibinfo {author} {\bibfnamefont
  {P.}~\bibnamefont {Xue}},\ }\bibfield  {title} {\enquote {\bibinfo {title}
  {Detecting topological invariants in nonunitary discrete-time quantum
  walks},}\ }\href@noop {} {\bibfield  {journal} {\bibinfo  {journal} {Phys.
  Rev. Lett.}\ }\textbf {\bibinfo {volume} {119}},\ \bibinfo {pages} {130501}
  (\bibinfo {year} {2017})}\BibitemShut {NoStop}%
\bibitem [{\citenamefont {Cardano}\ \emph {et~al.}(2017)\citenamefont
  {Cardano}, \citenamefont {D'Errico}, \citenamefont {Dauphin}, \citenamefont
  {Maffei}, \citenamefont {Piccirillo}, \citenamefont {De~Lisio}, \citenamefont
  {De~Filippis}, \citenamefont {Cataudella}, \citenamefont {Santamato},
  \citenamefont {Marrucci}, \citenamefont {Lewenstein},\ and\ \citenamefont
  {Massignan}}]{Cardano2017}%
  \BibitemOpen
  \bibfield  {author} {\bibinfo {author} {\bibfnamefont {F.}~\bibnamefont
  {Cardano}}, \bibinfo {author} {\bibfnamefont {A.}~\bibnamefont {D'Errico}},
  \bibinfo {author} {\bibfnamefont {A.}~\bibnamefont {Dauphin}}, \bibinfo
  {author} {\bibfnamefont {M.}~\bibnamefont {Maffei}}, \bibinfo {author}
  {\bibfnamefont {B.}~\bibnamefont {Piccirillo}}, \bibinfo {author}
  {\bibfnamefont {C.}~\bibnamefont {De~Lisio}}, \bibinfo {author}
  {\bibfnamefont {G.}~\bibnamefont {De~Filippis}}, \bibinfo {author}
  {\bibfnamefont {V.}~\bibnamefont {Cataudella}}, \bibinfo {author}
  {\bibfnamefont {E.}~\bibnamefont {Santamato}}, \bibinfo {author}
  {\bibfnamefont {L.}~\bibnamefont {Marrucci}}, \bibinfo {author}
  {\bibfnamefont {M.}~\bibnamefont {Lewenstein}}, \ and\ \bibinfo {author}
  {\bibfnamefont {P.}~\bibnamefont {Massignan}},\ }\bibfield  {title} {\enquote
  {\bibinfo {title} {Detection of {Z}ak phases and topological invariants in a
  chiral quantum walk of twisted photons},}\ }\href@noop {} {\bibfield
  {journal} {\bibinfo  {journal} {Nat. Commun.}\ }\textbf {\bibinfo {volume}
  {8}},\ \bibinfo {pages} {15516} (\bibinfo {year} {2017})}\BibitemShut
  {NoStop}%
\bibitem [{\citenamefont {Barkhofen}\ \emph {et~al.}(2017)\citenamefont
  {Barkhofen}, \citenamefont {Nitsche}, \citenamefont {Elster}, \citenamefont
  {Lorz}, \citenamefont {Gabris}, \citenamefont {Jex},\ and\ \citenamefont
  {Silberhorn}}]{silberhorn16}%
  \BibitemOpen
  \bibfield  {author} {\bibinfo {author} {\bibfnamefont {S.}~\bibnamefont
  {Barkhofen}}, \bibinfo {author} {\bibfnamefont {T.}~\bibnamefont {Nitsche}},
  \bibinfo {author} {\bibfnamefont {F.}~\bibnamefont {Elster}}, \bibinfo
  {author} {\bibfnamefont {L.}~\bibnamefont {Lorz}}, \bibinfo {author}
  {\bibfnamefont {A.}~\bibnamefont {Gabris}}, \bibinfo {author} {\bibfnamefont
  {I.}~\bibnamefont {Jex}}, \ and\ \bibinfo {author} {\bibfnamefont
  {C.}~\bibnamefont {Silberhorn}},\ }\bibfield  {title} {\enquote {\bibinfo
  {title} {Measuring topological invariants and protected bound states in
  disordered discrete time quantum walks},}\ }\href@noop {} {\bibfield
  {journal} {\bibinfo  {journal} {Phys. Rev. A}\ }\textbf {\bibinfo {volume}
  {96}},\ \bibinfo {pages} {033846} (\bibinfo {year} {2017})}\BibitemShut
  {NoStop}%
\bibitem [{\citenamefont {Zeuner}\ \emph {et~al.}(2015)\citenamefont {Zeuner},
  \citenamefont {Rechtsman}, \citenamefont {Plotnik}, \citenamefont {Lumer},
  \citenamefont {Nolte}, \citenamefont {Rudner}, \citenamefont {Segev},\ and\
  \citenamefont {Szameit}}]{Zeunerprl}%
  \BibitemOpen
  \bibfield  {author} {\bibinfo {author} {\bibfnamefont {J.~M.}\ \bibnamefont
  {Zeuner}}, \bibinfo {author} {\bibfnamefont {M.~C.}\ \bibnamefont
  {Rechtsman}}, \bibinfo {author} {\bibfnamefont {Y.}~\bibnamefont {Plotnik}},
  \bibinfo {author} {\bibfnamefont {Y.}~\bibnamefont {Lumer}}, \bibinfo
  {author} {\bibfnamefont {S.}~\bibnamefont {Nolte}}, \bibinfo {author}
  {\bibfnamefont {M.~S.}\ \bibnamefont {Rudner}}, \bibinfo {author}
  {\bibfnamefont {M.}~\bibnamefont {Segev}}, \ and\ \bibinfo {author}
  {\bibfnamefont {A.}~\bibnamefont {Szameit}},\ }\bibfield  {title} {\enquote
  {\bibinfo {title} {Observation of a topological transition in the bulk of a
  non-{H}ermitian system},}\ }\href@noop {} {\bibfield  {journal} {\bibinfo
  {journal} {Phys. Rev. Lett.}\ }\textbf {\bibinfo {volume} {115}},\ \bibinfo
  {pages} {040402} (\bibinfo {year} {2015})}\BibitemShut {NoStop}%
\bibitem [{\citenamefont {Rudner}\ \emph {et~al.}(2016)\citenamefont {Rudner},
  \citenamefont {Levin},\ and\ \citenamefont {Levitov}}]{RLL16}%
  \BibitemOpen
  \bibfield  {author} {\bibinfo {author} {\bibfnamefont {M.~S.}\ \bibnamefont
  {Rudner}}, \bibinfo {author} {\bibfnamefont {M.}~\bibnamefont {Levin}}, \
  and\ \bibinfo {author} {\bibfnamefont {L.~S.}\ \bibnamefont {Levitov}},\
  }\bibfield  {title} {\enquote {\bibinfo {title} {Survival, decay, and
  topological protection in non-{H}ermitian quantum transport},}\ }\href@noop
  {} {\bibfield  {journal} {\bibinfo  {journal} {preprint arXiv:1605.07652}\ }
  (\bibinfo {year} {2016})}\BibitemShut {NoStop}%
\bibitem [{\citenamefont {Asb{\'o}th}\ and\ \citenamefont
  {Obuse}(2013)}]{AO13}%
  \BibitemOpen
  \bibfield  {author} {\bibinfo {author} {\bibfnamefont {J.~K.}\ \bibnamefont
  {Asb{\'o}th}}\ and\ \bibinfo {author} {\bibfnamefont {H.}~\bibnamefont
  {Obuse}},\ }\bibfield  {title} {\enquote {\bibinfo {title} {Bulk-boundary
  correspondence for chiral symmetric quantum walks},}\ }\href@noop {}
  {\bibfield  {journal} {\bibinfo  {journal} {Phys. Rev. B}\ }\textbf {\bibinfo
  {volume} {88}},\ \bibinfo {pages} {121406} (\bibinfo {year}
  {2013})}\BibitemShut {NoStop}%
\bibitem [{\citenamefont {Kim}\ \emph {et~al.}(2016)\citenamefont {Kim},
  \citenamefont {Ken}, \citenamefont {Kawakami},\ and\ \citenamefont
  {Obuse}}]{KMKO16}%
  \BibitemOpen
  \bibfield  {author} {\bibinfo {author} {\bibfnamefont {D.}~\bibnamefont
  {Kim}}, \bibinfo {author} {\bibfnamefont {M.}~\bibnamefont {Ken}}, \bibinfo
  {author} {\bibfnamefont {N.}~\bibnamefont {Kawakami}}, \ and\ \bibinfo
  {author} {\bibfnamefont {H.}~\bibnamefont {Obuse}},\ }\bibfield  {title}
  {\enquote {\bibinfo {title} {Floquet topological phases driven by {P}{T}
  symmetric nonunitary time evolution},}\ }\href@noop {} {\bibfield  {journal}
  {\bibinfo  {journal} {preprint arXiv:1609.09650}\ } (\bibinfo {year}
  {2016})}\BibitemShut {NoStop}%
\bibitem [{\citenamefont {Jiang}\ \emph {et~al.}(2012)\citenamefont {Jiang},
  \citenamefont {Qiao}, \citenamefont {Liu},\ and\ \citenamefont
  {Niu}}]{Chern1}%
  \BibitemOpen
  \bibfield  {author} {\bibinfo {author} {\bibfnamefont {H.}~\bibnamefont
  {Jiang}}, \bibinfo {author} {\bibfnamefont {Z.~H.}\ \bibnamefont {Qiao}},
  \bibinfo {author} {\bibfnamefont {H.~W.}\ \bibnamefont {Liu}}, \ and\
  \bibinfo {author} {\bibfnamefont {Q.}~\bibnamefont {Niu}},\ }\bibfield
  {title} {\enquote {\bibinfo {title} {Quantum anomalous {H}all effect with
  tunable {C}hern number in magnetic topological insulator film},}\ }\href@noop
  {} {\bibfield  {journal} {\bibinfo  {journal} {Phys. Rev. B}\ }\textbf
  {\bibinfo {volume} {85}},\ \bibinfo {pages} {045445} (\bibinfo {year}
  {2012})}\BibitemShut {NoStop}%
\bibitem [{\citenamefont {Fang}\ \emph {et~al.}(2014)\citenamefont {Fang},
  \citenamefont {Gilbert},\ and\ \citenamefont {Bernevig}}]{Chern2}%
  \BibitemOpen
  \bibfield  {author} {\bibinfo {author} {\bibfnamefont {C.}~\bibnamefont
  {Fang}}, \bibinfo {author} {\bibfnamefont {M.~J.}\ \bibnamefont {Gilbert}}, \
  and\ \bibinfo {author} {\bibfnamefont {B.~A.}\ \bibnamefont {Bernevig}},\
  }\bibfield  {title} {\enquote {\bibinfo {title} {Large-{C}hern-number quantum
  anomalous {H}all effect in thin-film topological crystalline insulators},}\
  }\href@noop {} {\bibfield  {journal} {\bibinfo  {journal} {Phys. Rev. Lett.}\
  }\textbf {\bibinfo {volume} {112}},\ \bibinfo {pages} {046801} (\bibinfo
  {year} {2014})}\BibitemShut {NoStop}%
\bibitem [{\citenamefont {Skirlo}\ \emph {et~al.}(2014)\citenamefont {Skirlo},
  \citenamefont {Lu},\ and\ \citenamefont {Solja{\v{c}}i{\'c}}}]{Chern3}%
  \BibitemOpen
  \bibfield  {author} {\bibinfo {author} {\bibfnamefont {S.~A.}\ \bibnamefont
  {Skirlo}}, \bibinfo {author} {\bibfnamefont {L.}~\bibnamefont {Lu}}, \ and\
  \bibinfo {author} {\bibfnamefont {M.}~\bibnamefont {Solja{\v{c}}i{\'c}}},\
  }\bibfield  {title} {\enquote {\bibinfo {title} {Multimode one-way waveguides
  of large {C}hern numbers},}\ }\href@noop {} {\bibfield  {journal} {\bibinfo
  {journal} {Phys. Rev. Lett.}\ }\textbf {\bibinfo {volume} {113}},\ \bibinfo
  {pages} {113904} (\bibinfo {year} {2014})}\BibitemShut {NoStop}%
\bibitem [{\citenamefont {Rudner}\ and\ \citenamefont {Levitov}(2009)}]{RL09}%
  \BibitemOpen
  \bibfield  {author} {\bibinfo {author} {\bibfnamefont {M.~S.}\ \bibnamefont
  {Rudner}}\ and\ \bibinfo {author} {\bibfnamefont {L.~S.}\ \bibnamefont
  {Levitov}},\ }\bibfield  {title} {\enquote {\bibinfo {title} {Topological
  transition in a non-{H}ermitian quantum walk},}\ }\href@noop {} {\bibfield
  {journal} {\bibinfo  {journal} {Phys. Rev. Lett.}\ }\textbf {\bibinfo
  {volume} {102}},\ \bibinfo {pages} {065703} (\bibinfo {year}
  {2009})}\BibitemShut {NoStop}%
\bibitem [{\citenamefont {Rakovszky}\ \emph {et~al.}(2017)\citenamefont
  {Rakovszky}, \citenamefont {Asb{\'o}th},\ and\ \citenamefont
  {Alberti}}]{RAA17}%
  \BibitemOpen
  \bibfield  {author} {\bibinfo {author} {\bibfnamefont {T.}~\bibnamefont
  {Rakovszky}}, \bibinfo {author} {\bibfnamefont {J.~K.}\ \bibnamefont
  {Asb{\'o}th}}, \ and\ \bibinfo {author} {\bibfnamefont {A.}~\bibnamefont
  {Alberti}},\ }\bibfield  {title} {\enquote {\bibinfo {title} {Detecting
  topological invariants in chiral symmetric insulators via losses},}\
  }\href@noop {} {\bibfield  {journal} {\bibinfo  {journal} {Physical Rev. B}\
  }\textbf {\bibinfo {volume} {95}},\ \bibinfo {pages} {201407} (\bibinfo
  {year} {2017})}\BibitemShut {NoStop}%
\bibitem [{\citenamefont {Kitagawa}\ \emph {et~al.}(2010)\citenamefont
  {Kitagawa}, \citenamefont {Rudner}, \citenamefont {Berg},\ and\ \citenamefont
  {Demler}}]{CSPRA}%
  \BibitemOpen
  \bibfield  {author} {\bibinfo {author} {\bibfnamefont {T.}~\bibnamefont
  {Kitagawa}}, \bibinfo {author} {\bibfnamefont {M.~S.}\ \bibnamefont
  {Rudner}}, \bibinfo {author} {\bibfnamefont {E.}~\bibnamefont {Berg}}, \ and\
  \bibinfo {author} {\bibfnamefont {E.}~\bibnamefont {Demler}},\ }\bibfield
  {title} {\enquote {\bibinfo {title} {Exploring topological phases with
  quantum walks},}\ }\href@noop {} {\bibfield  {journal} {\bibinfo  {journal}
  {Phys. Rev. A}\ }\textbf {\bibinfo {volume} {82}},\ \bibinfo {pages} {033429}
  (\bibinfo {year} {2010})}\BibitemShut {NoStop}%
\bibitem [{sup()}]{supp}%
  \BibitemOpen
  \href@noop {} {}\bibinfo {note} {In 
  provide details on the experimental implementation, the defnition of
  topological invariants, the numerical confrmation of bulk-boundary
  correspondence, statistical moments of quantum walks, the pseudo-unitarity of
  the Floquet operators, as well as the robustness of topological
  properties.}\BibitemShut {Stop}%
\bibitem [{\citenamefont {Mostafazadeh}(2002{\natexlab{a}})}]{M2002a}%
  \BibitemOpen
  \bibfield  {author} {\bibinfo {author} {\bibfnamefont {A.}~\bibnamefont
  {Mostafazadeh}},\ }\bibfield  {title} {\enquote {\bibinfo {title}
  {Pseudo-{H}ermiticity versus {P}{T} symmetry: the necessary condition for the
  reality of the spectrumof a non-{H}ermitian {H}amiltonian},}\ }\href@noop {}
  {\bibfield  {journal} {\bibinfo  {journal} {J. Math. Phys.}\ }\textbf
  {\bibinfo {volume} {43}},\ \bibinfo {pages} {205--214} (\bibinfo {year}
  {2002}{\natexlab{a}})}\BibitemShut {NoStop}%
\bibitem [{\citenamefont {Mostafazadeh}(2002{\natexlab{b}})}]{M2002b}%
  \BibitemOpen
  \bibfield  {author} {\bibinfo {author} {\bibfnamefont {A.}~\bibnamefont
  {Mostafazadeh}},\ }\bibfield  {title} {\enquote {\bibinfo {title}
  {Pseudo-{H}ermiticity versus {P}{T}-symmetry. {I}{I}. a complete
  characterization of non-{H}ermitian {H}amiltonians with a real spectrum},}\
  }\href@noop {} {\bibfield  {journal} {\bibinfo  {journal} {J. Math. Phys.}\
  }\textbf {\bibinfo {volume} {43}},\ \bibinfo {pages} {2814--2816} (\bibinfo
  {year} {2002}{\natexlab{b}})}\BibitemShut {NoStop}%
\bibitem [{\citenamefont {Mostafazadeh}(2004)}]{M2004}%
  \BibitemOpen
  \bibfield  {author} {\bibinfo {author} {\bibfnamefont {A.}~\bibnamefont
  {Mostafazadeh}},\ }\bibfield  {title} {\enquote {\bibinfo {title}
  {Pseudounitary operators and pseudounitary quantum dynamics},}\ }\href@noop
  {} {\bibfield  {journal} {\bibinfo  {journal} {J. Math. Phys.}\ }\textbf
  {\bibinfo {volume} {45}},\ \bibinfo {pages} {932--946} (\bibinfo {year}
  {2004})}\BibitemShut {NoStop}%
\bibitem [{\citenamefont {Regensburger}\ \emph {et~al.}(2012)\citenamefont
  {Regensburger}, \citenamefont {Bersch}, \citenamefont {Miri}, \citenamefont
  {Onishchukov}, \citenamefont {Christodoulides},\ and\ \citenamefont
  {Peschel}}]{PTsymm1}%
  \BibitemOpen
  \bibfield  {author} {\bibinfo {author} {\bibfnamefont {A.}~\bibnamefont
  {Regensburger}}, \bibinfo {author} {\bibfnamefont {C.}~\bibnamefont
  {Bersch}}, \bibinfo {author} {\bibfnamefont {M.~A.}\ \bibnamefont {Miri}},
  \bibinfo {author} {\bibfnamefont {G.}~\bibnamefont {Onishchukov}}, \bibinfo
  {author} {\bibfnamefont {D.~N.}\ \bibnamefont {Christodoulides}}, \ and\
  \bibinfo {author} {\bibfnamefont {U.}~\bibnamefont {Peschel}},\ }\bibfield
  {title} {\enquote {\bibinfo {title} {Parity-time synthetic photonic
  lattices},}\ }\href@noop {} {\bibfield  {journal} {\bibinfo  {journal}
  {Nature}\ }\textbf {\bibinfo {volume} {488}},\ \bibinfo {pages} {167}
  (\bibinfo {year} {2012})}\BibitemShut {NoStop}%
\bibitem [{\citenamefont {Mochizuki}\ \emph {et~al.}(2016)\citenamefont
  {Mochizuki}, \citenamefont {Kim},\ and\ \citenamefont {Obuse}}]{PTbroken}%
  \BibitemOpen
  \bibfield  {author} {\bibinfo {author} {\bibfnamefont {K.}~\bibnamefont
  {Mochizuki}}, \bibinfo {author} {\bibfnamefont {D.}~\bibnamefont {Kim}}, \
  and\ \bibinfo {author} {\bibfnamefont {H.}~\bibnamefont {Obuse}},\ }\bibfield
   {title} {\enquote {\bibinfo {title} {Explicit definition of {P}{T} symmetry
  for nonunitary quantum walks with gain and loss},}\ }\href@noop {} {\bibfield
   {journal} {\bibinfo  {journal} {Phys. Rev. A}\ }\textbf {\bibinfo {volume}
  {93}},\ \bibinfo {pages} {062116} (\bibinfo {year} {2016})}\BibitemShut
  {NoStop}%
\bibitem [{\citenamefont {Schreiber}\ \emph {et~al.}(2010)\citenamefont
  {Schreiber}, \citenamefont {Cassemiro}, \citenamefont {Poto{\v{c}}ek},
  \citenamefont {G{\'a}bris}, \citenamefont {Mosley}, \citenamefont
  {Andersson}, \citenamefont {Jex},\ and\ \citenamefont {Silberhorn}}]{SCP+10}%
  \BibitemOpen
  \bibfield  {author} {\bibinfo {author} {\bibfnamefont {A.}~\bibnamefont
  {Schreiber}}, \bibinfo {author} {\bibfnamefont {K.~N.}\ \bibnamefont
  {Cassemiro}}, \bibinfo {author} {\bibfnamefont {V.}~\bibnamefont
  {Poto{\v{c}}ek}}, \bibinfo {author} {\bibfnamefont {A.}~\bibnamefont
  {G{\'a}bris}}, \bibinfo {author} {\bibfnamefont {P.~J.}\ \bibnamefont
  {Mosley}}, \bibinfo {author} {\bibfnamefont {E.}~\bibnamefont {Andersson}},
  \bibinfo {author} {\bibfnamefont {I.}~\bibnamefont {Jex}}, \ and\ \bibinfo
  {author} {\bibfnamefont {C.}~\bibnamefont {Silberhorn}},\ }\bibfield  {title}
  {\enquote {\bibinfo {title} {Photons walking the line: a quantum walk with
  adjustable coin operations},}\ }\href@noop {} {\bibfield  {journal} {\bibinfo
   {journal} {Phys. Rev. Lett.}\ }\textbf {\bibinfo {volume} {104}},\ \bibinfo
  {pages} {050502} (\bibinfo {year} {2010})}\BibitemShut {NoStop}%
\bibitem [{\citenamefont {Schreiber}\ \emph {et~al.}(2011)\citenamefont
  {Schreiber}, \citenamefont {Cassemiro}, \citenamefont {Poto{\v{c}}ek},
  \citenamefont {G{\'a}bris}, \citenamefont {Jex},\ and\ \citenamefont
  {Silberhorn}}]{SCP+11}%
  \BibitemOpen
  \bibfield  {author} {\bibinfo {author} {\bibfnamefont {A.}~\bibnamefont
  {Schreiber}}, \bibinfo {author} {\bibfnamefont {K.~N.}\ \bibnamefont
  {Cassemiro}}, \bibinfo {author} {\bibfnamefont {V.}~\bibnamefont
  {Poto{\v{c}}ek}}, \bibinfo {author} {\bibfnamefont {A.}~\bibnamefont
  {G{\'a}bris}}, \bibinfo {author} {\bibfnamefont {I.}~\bibnamefont {Jex}}, \
  and\ \bibinfo {author} {\bibfnamefont {C.}~\bibnamefont {Silberhorn}},\
  }\bibfield  {title} {\enquote {\bibinfo {title} {Decoherence and disorder in
  quantum walks: from ballistic spread to localization},}\ }\href@noop {}
  {\bibfield  {journal} {\bibinfo  {journal} {Phys. Rev. Lett.}\ }\textbf
  {\bibinfo {volume} {106}},\ \bibinfo {pages} {180403} (\bibinfo {year}
  {2011})}\BibitemShut {NoStop}%
\bibitem [{\citenamefont {Schreiber}\ \emph {et~al.}(2012)\citenamefont
  {Schreiber}, \citenamefont {G{\'a}bris}, \citenamefont {Rohde}, \citenamefont
  {Laiho}, \citenamefont {{\v{S}}tefa{\v{n}}{\'a}k}, \citenamefont
  {Poto{\v{c}}ek}, \citenamefont {Hamilton}, \citenamefont {Jex},\ and\
  \citenamefont {Silberhorn}}]{SGP+12}%
  \BibitemOpen
  \bibfield  {author} {\bibinfo {author} {\bibfnamefont {A.}~\bibnamefont
  {Schreiber}}, \bibinfo {author} {\bibfnamefont {A.}~\bibnamefont
  {G{\'a}bris}}, \bibinfo {author} {\bibfnamefont {P.~P.}\ \bibnamefont
  {Rohde}}, \bibinfo {author} {\bibfnamefont {K.}~\bibnamefont {Laiho}},
  \bibinfo {author} {\bibfnamefont {M.}~\bibnamefont
  {{\v{S}}tefa{\v{n}}{\'a}k}}, \bibinfo {author} {\bibfnamefont
  {V.}~\bibnamefont {Poto{\v{c}}ek}}, \bibinfo {author} {\bibfnamefont
  {C.}~\bibnamefont {Hamilton}}, \bibinfo {author} {\bibfnamefont
  {I.}~\bibnamefont {Jex}}, \ and\ \bibinfo {author} {\bibfnamefont
  {C.}~\bibnamefont {Silberhorn}},\ }\bibfield  {title} {\enquote {\bibinfo
  {title} {A 2{D} quantum walk simulation of two-particle dynamics},}\
  }\href@noop {} {\bibfield  {journal} {\bibinfo  {journal} {Science}\ }\textbf
  {\bibinfo {volume} {336}},\ \bibinfo {pages} {55--58} (\bibinfo {year}
  {2012})}\BibitemShut {NoStop}%
\end{thebibliography}%

\clearpage
\begin{widetext}
\appendix

In Appendix, we provide details on the experimental realization, the definition of topological invariants, choice of topological Floquet operators, the numerical confirmation of bulk-boundary correspondence, statistical moments of the unitary and non-unitary quantum walks (QWs), the pseudo-unitarity of the Floquet operators governing the quantum-walk dynamics, the average chiral displacement, topological edge states,
as well as the robustness of Floquet topological phases (FTPs) with large winding numbers against disorder.

\section{Experimental Realization of Multi-Step Non-unitary Quantum Walks}
At the start of a single-photon QW, a pair of photons is generated via type-I spontaneous parametric downconversion, with one photon serving as a trigger. The other photon is projected into the state $\ket{+}$ with a polarizing beamsplitter (PBS) and a half-wave plate (HWP) heralded by the trigger photon, and is then sent to the quantum-walk interferometric setup. The coin operator $R(\theta)=\one_\text{w}\otimes \text{e}^{-\text{i}\theta\sigma_y}$, the conditional position shift operator $S=\sum_x\left(\ket{x-1}\bra{x}\otimes\ket{H}\bra{H}+\ket{x+1}\bra{x}\otimes\ket{V}\bra{V}\right)$, and the partial measurement operator $M_{\text{e}}$ are realized using similar methods in~\cite{pxprl}. Here, $\sigma_y$ is one of the standard Pauli operators.

Losses are used to detect winding numbers in our experiment and can be controlled by the transmissivity of the partial polarizing beam splitter. Each pair of beam displacers forms an interferometer and their misalignment gives rise to pure dephasing, which is the major form of decoherence in the system. Furthermore, the surfaces of the beam displacers are not strictly smooth due to manufacturing inaccuracy. These should give rise to position-dependent dephasing throughout the QW. However, the dephasing caused by misalignment between beam displacers and imperfectness of the surface of the beam displacer can be compensated experimentally. Ideally, losses and misalignment of beam displacers do not limit the number of steps. The limitation on the number of steps depends on the size of the clear aperture of the beam displacer, which can be relaxed at the cost of beam displacers with larger clear apertures. Therefore, whereas we demonstrate that, by choosing the proper parameters, as few as four (three) steps are enough to have a clear detection of higher winding numbers, four (three) steps are not the limit of our experimental setups.

\section{Topological invariants for multi-step non-unitary quantum walks.}
In this section, we define winding numbers for multi-step non-unitary QWs and discuss their relation with the topological edge states.

We write the Floquet operator
\begin{equation}
\widetilde{U}'_3 = n_0\sigma_0-\text{i}n_1\sigma_x-\text{i}n_2\sigma_y-\text{i}n_3\sigma_z
\end{equation}
in momentum space. We then define a new vector $\bf{h}:=\frac{1}{ \|\text{Re}\left(\bf{n}\right)\|}\mathrm{Re}(\bf{n})$, with $\bf{n}=(n_1,n_2,n_3)^T$.
As $h_1=\frac{1}{ \|\text{Re}\left(\bf{n}\right)\|}\mathrm{Re}(n_1)=0$ for all $k$, the topological invariant for the non-unitary QW is
\begin{equation}
\nu':=-\frac{1}{2\pi}\oint \text{d}k\left(\bf{h}\times\frac{\partial{\bf{h}}}{\partial{k}}\right)_1.
\label{eqn:nu31}
\end{equation}
Following a similar procedure, we define the winding number $\nu''$ for the Floquet operator $\widetilde{U}''_3$. Note that the winding number Eq.~(\ref{eqn:nu31}) is defined through the spinor eigen-vectors of $\widetilde{U}'_3$. This is equivalent to the definition through the spin eigen-vectors of the corresponding effective Hamiltonian $H'^{(3)}_{\rm eff}$.

In the unitary limit with $p=0$, $n_1$ becomes zero and the Floquet operators $\widetilde{U}'_3$ and $\widetilde{U}''_3$ manifestly satisfy chiral symmetry, with the chiral symmetry operator being $\Gamma=\sigma_x$. We define~\cite{AO13}
\begin{equation}
(\nu_0,\nu_{\pi}):=\left(\frac{\nu'+\nu''}{2},\frac{\nu'-\nu''}{2}\right),
\end{equation}
which are directly related to edge states at the boundaries with quasienergies $0$ and $\pi$, respectively. Specifically, the number of edge states with quasienergy $0$ ($\pi$) should be equal to the difference in the winding numbers $\nu_0$ ($\nu_{\pi}$) on either side of the boundary. For the non-unitary QW ($p>0$), the Floquet operators no longer possess chiral symmetry, and the bulk-boundary correspondence between the bulk winding numbers and the topological edge states needs to be confirmed. We have checked numerically that the topological invariants $\nu_0$ and $\nu_{\pi}$ are related to localized topological edge states with the real-parts of quasienergies at $0$ and $\pi$, respectively.

Following the same recipe, we define the winding numbers for the four-step non-unitary QW governed by $\widetilde{U}'_4$ and $\widetilde{U}''_4$, respectively. We note that the Floquet operators, the coin parameters in particular, are chosen such that the operators possess chiral symmetries in the unitary limit ($p=0$). We have also checked numerically that the bulk-boundary correspondence holds for the four-step non-unitary QW as well.

\section{Choice of topological Floquet operators}
As we have discussed previously, $\widetilde{U}'_3$ and $\widetilde{U}'_4$ are topologically non-trivial so long as they possess chiral symmetry in the unitary limit. This allows us freedom in the design of multi-step QWs. As an example, we consider four-step non-unitary QWs under the Floquet operators
\begin{align}
\widetilde{W}'_4&=MR\left(\frac{\theta_1}{2}\right)SR(\theta_2)SSR(\theta_2)SR\left(\frac{\theta_1}{2}\right).\\
\widetilde{W}''_4&= MSR(\theta_2)SR\left(\frac{\theta_1}{2}\right)R\left(\frac{\theta_1}{2}\right)SR(\theta_2)S.
\label{eq:U4}
\end{align}
The corresponding phase diagram is shown in Fig.~\ref{fig:phase}, which is richer than that of the four-step QW in the main text. We then calculated average displacements under $\widetilde{W}'_4$ with different loss parameters, the results are shown in the middle and right panels of Fig.~\ref{fig:phase}. Whereas under our parameters, the average displacements have not yet converged at four time steps, the topological nature of $\widetilde{W}'_4$ is revealed by the quantized average displacements at long times.

\begin{figure}
\includegraphics[width=0.85\textwidth]{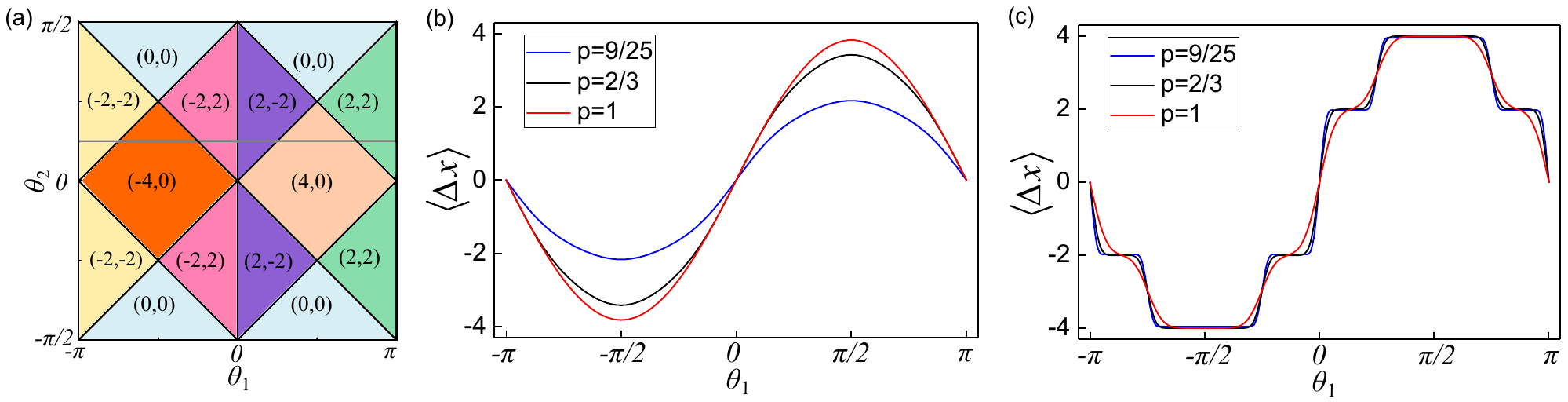}
\caption{ Left: phase diagram for new four-step non-unitary QWs in terms of the topological invariants $(\nu',\nu'')$. $(\nu',\nu'')$ are calculated from the Floquet operators $\widetilde{W}'_4$ and $\widetilde{W}''_4$ respectively. Middle: average displacements under $\widetilde{W}'_4$ for $3$ time steps. Coin parameters vary along the gray line $\theta_2=\pi/8$ in the phase diagram. The coloured solid curve indicates numerical simulations for different loss parameter. The blue solid lines indicate $p=9/25$, black solid lines indicate $p=2/3$ and red solid lines indicate $p=1$. The walker starts from $x=0$ and the initial coin state was chosen as $|+\rangle$. Right:  average displacements under $\widetilde{W}'_4$ for $20$ time steps. Other parameters are the same as those of the middle panel.}
\label{fig:phase}
\end{figure}

\section{Winding numbers and the bulk-boundary correspondence}

In this section, we numerically confirm the bulk-boundary correspondence for the multi-step non-unitary QW. As shown in the Methods, from the detected winding numbers $(\nu',\nu'')$, we construct the topological invariants $(\nu_0,\nu_{\pi})=\left(\frac{\nu'+\nu''}{2},\frac{\nu'-\nu''}{2}\right)$. We numerically confirm that the number of localized topological edge states with the real-part of quasienergy at $0$ ($\pi$) equals the difference in the winding numbers $\nu_0$ ($\nu_{\pi}$) on either side of the boundary. For brevity, we use the three-step non-unitary QW as an example. The case with the four-step non-unitary QW is similar.

We consider an inhomogeneous three-step QW on a lattice with $401$ sites and with a periodic boundary condition. The non-unitary QW is governed by the Floquet operator $\widetilde{U}_3'$ with $p=9/25$. We introduce two boundaries near $x=0$ and $x=\pm 200$, with $(\theta^{\rm L}_1,\theta^{\rm L}_2)=(\pi/4,-\pi/4)$ for $-200\leq x<0$ and $(\theta^{\rm R}_1,\theta^{\rm R}_2)=(\pi/2,0)$ for $0\leq x\leq 200$.
According to the phase diagram in Fig.~\ref{displacement3}(a) in the main text, the winding numbers for $-200\leq x<0$ is $(\nu',\nu'')=(1,0)$, and those for $0\leq x\leq 200$ is $(\nu',\nu'')=(3,0)$. Therefore, we have $(\nu^{L}_0,\nu^{L}_{\pi})=(\frac{1}{2},\frac{1}{2})$ for $-200\leq x<0$, and $(\nu^R_0,\nu^R_{\pi})=(\frac{3}{2},\frac{3}{2})$ for $0\leq x\leq 200$.

At both boundaries, the differences between the winding numbers are $\delta\nu_0=\left|\nu_0^L-\nu_0^R\right|=1$ and $\delta\nu_{\pi}=\left|\nu_{\pi}^L-\nu_{\pi}^R\right|=1$. This should correspond to a pair of topological edge states at each edge, with the real-part of their quasienergies at $0$ and $\pi$, respectively. In the following, we confirm this expectation by numerically calculating the quasienergy spectrum.

We define the effective Hamiltonian $\widetilde{U}_3'=\exp(-\text{i} H_{\rm eff})$. The quasienergy $\epsilon$ is defined as
\begin{align}
\widetilde{U}_3'|\psi_\lambda\rangle=\lambda|\psi_\lambda\rangle,\quad \lambda=e^{-\text{i}\epsilon},
\end{align}
where $|\psi_\lambda\rangle$ is the eigenstate of $\widetilde{U}_3'$ and $H_{\rm eff}$. In Fig.~\ref{fig:s0}(a), we plot the eigen-spectrum of $\lambda$ on the complex plane. Whereas the blue dots are the bulk states, the red (B and C) and the black (A and D) dots appearing on the real axis correspond to localized edge states at the two boundaries near $x=0$ and $x=\pm 200$, respectively.
Localization of the edges states is confirmed by plotting the probability distribution $P_x=\langle \psi_{\lambda}|x\rangle\langle x|\otimes\one_{\text{c}}|\psi_{\lambda}\rangle$ of the edge states (A, B, C, and D), as illustrated in Fig.~\ref{fig:s0}(b).

For comparison, we have shown typical spatial distributions of the bulk states (E and F), which are indeed extended in space. Importantly, near $x=0$, there exist two localized edge states with identical spatial distributions, which correspond to the red dots (B and C) in Fig.~\ref{fig:s0}(a). The real parts of the corresponding quasienergie $\epsilon$ are given by $\pi$ (B) and $0$ (C), respectively. The case at the boundary near $x=\pm 200$ is similar. For comparison, we have also shown typical spatial distributions of the bulk states (E and F), which are indeed extended in space.
This confirms the bulk-boundary correspondence as discussed in the previous paragraph. We have checked that such a bulk-boundary correspondence works for other choices of coin parameters throughout the phase diagram in Fig.~\ref{displacement3}(a) in the main text.

\section{Statistical moments of quantum walks}
In this section, we examine the statistical moments of both the unitary and the non-unitary QWs. Consider a general homogeneous QW driven by the Floquet operator $U = n_0\sigma_0-\text{i}\bf{n}\cdot\bf{\sigma}$. Assuming the walker starts from $x=0$ at $t=0$, we write the initial state of the walker-coin system as $|\Psi_0\rangle=|x=0\rangle\otimes|\psi_0\rangle$, where $|\psi_0\rangle$ represents the coin state. At any given time step $t>0$, we have $|\Psi_t\rangle=U^t|\Psi_0\rangle$, and the probability of measuring the walker at position $x$ is
\begin{equation}
p(x,t)=\langle \Psi_t|x\rangle\langle x|\otimes\one_{\text{c}}|\Psi_t\rangle.
\end{equation}

The $j^{\text{th}}$ statistical moment of this distribution is given by $m_j(t)=\langle x^j\rangle_t=\sum_xx^jp(x,t)$. In particular, we write the second moment in the momentum space as
\begin{equation}
\begin{split}
m_2(t) = &~ \int^\pi_{-\pi}\frac{\text{d}k}{2\pi}\langle\psi_0|U^{\dag t}\left(-\text{i}\frac{\text{d}}{\text{d}k}\right)^2U^t|\psi_0\rangle.
\end{split}
\end{equation}

For the unitary QWs, we have $U=\cos E_k\sigma_0-\text{i}\sin E_k(\hat{\bf n}_k\cdot\bf{\sigma})$ and hence $U^t=\cos(E_kt)\sigma_0-\text{i}\sin(E_kt)(\hat{\bf{n}}_k\cdot{\bf \sigma})$, where $\hat{\bf{n}}_k={\bf n}_k/\sin E_k$ and $\cos E_k=n_0$. It is then straightforward to derive
\begin{equation}
\begin{split}
\frac{m_2(t)}{t^2} = &~ \int^\pi_{-\pi}\frac{\text{d}k}{2\pi}v^2_k+O\left(1/t^2\right),\\
\int^\pi_{-\pi}\frac{\text{d}k}{2\pi}v^2_k = &~\int^\pi_{-\pi}\frac{\text{d}k}{2\pi}\left(\frac{\text{d}E_k}{\text{d}k}\right)^2=\int^\pi_{-\pi}\frac{\text{d}k}{2\pi}\frac{1}{1-n^2_0}\left(\frac{\text{d}n_0}{\text{d}k}\right)^2,
\end{split}
\label{eqn:groupvelocity}
\end{equation}
where $v_k=\frac{\text{d}E_k}{\text{d}k}$ is the group velocity. At the topological phase boundary, the bulk gap closes at certain points in the momentum space, and the corresponding $n_0(k)$ at these momenta approaches zero. This gives rise to the slope discontinuity, as well as a peak structure of the second moment near the phase boundary~\cite{Cardano2016}.

\begin{figure}
\includegraphics[width=0.8\textwidth]{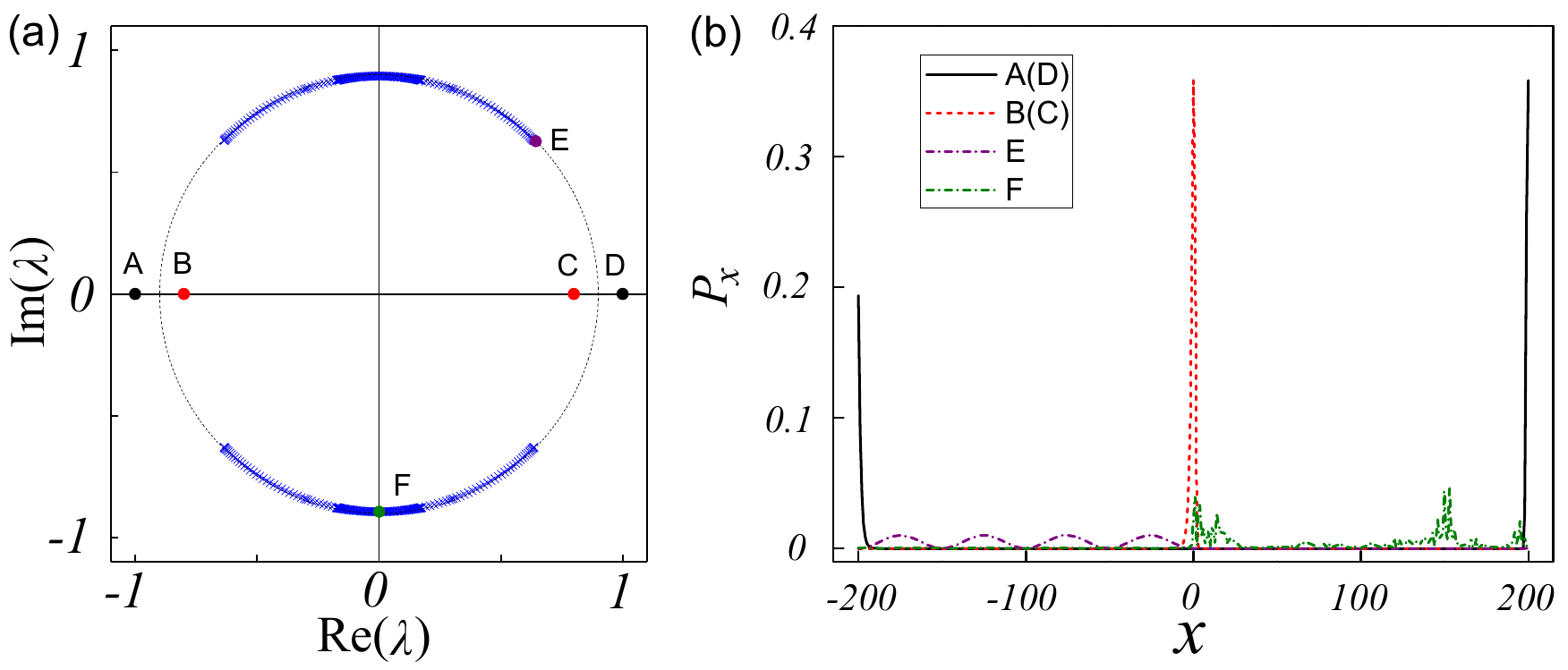}
\caption{(a) Eigen-spectrum of $\lambda$ on the complex plane. The blue dots correspond to bulk states, and the red (black) dots correspond to topological edge states located near $x=0$ and $x=\pm 200$, respectively. There are altogether $4$ edge states in the spectrum. (b) The spatial probability distribution of the edge states corresponding to the red and black dots in (a).}
\label{fig:s0}
\end{figure}

\begin{figure}
\includegraphics[width=0.8\textwidth]{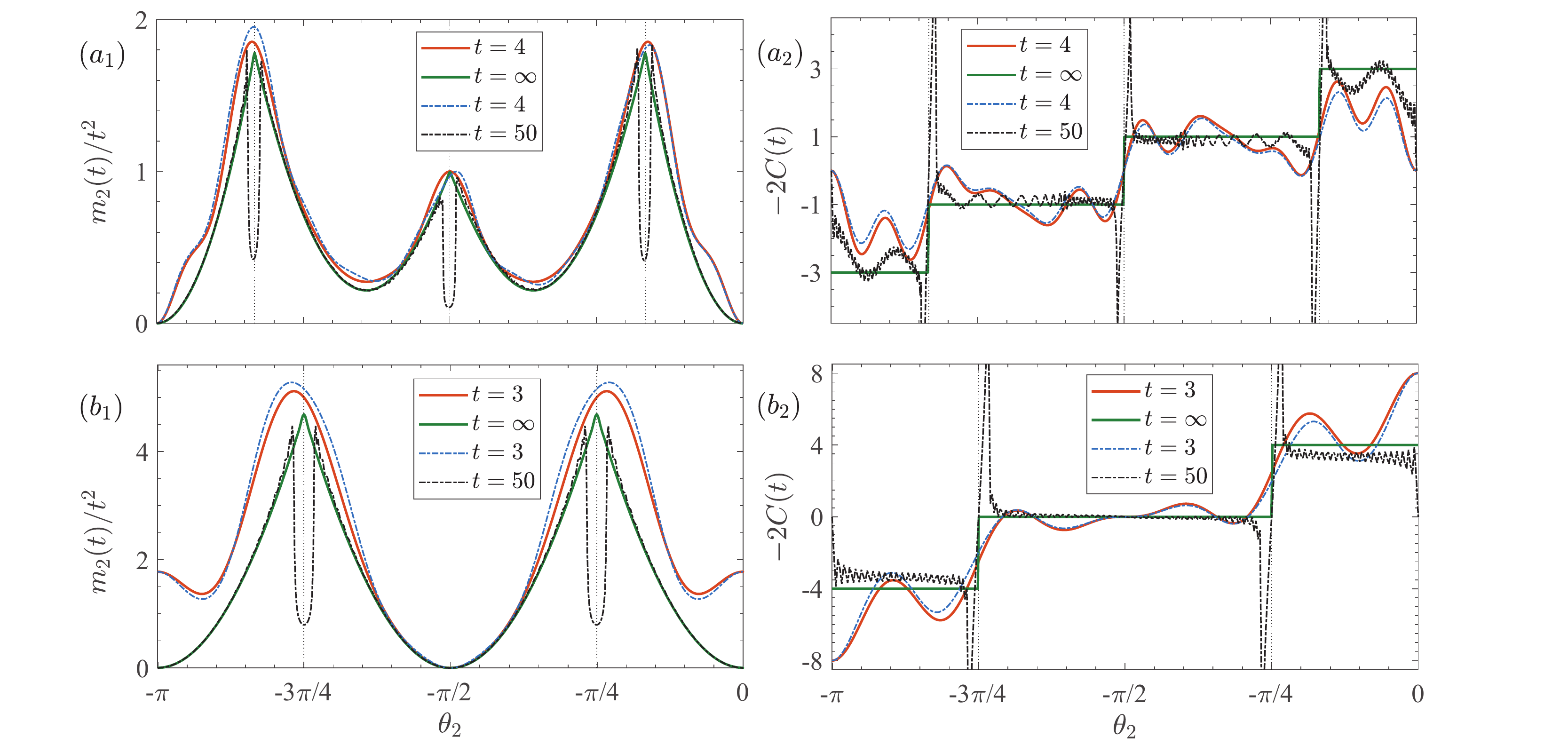}
\caption{Statistical moments $m_2(t)/t^2$ and scaled chiral displacements $-2C(t)$ of the walker position distribution for unitary (solid curves) and non-unitary (dotted-dashed curves), three-step (upper layer) and four-step (lower layer) QWs, governed by $U_3'$, $\widetilde{U}_3'$, $U_4'$ and $\widetilde{U}_4'$, respectively. The green solid curves in (a1) and (b1) indicate the analytical results of the second moment (see Eq.~(\ref{eqn:groupvelocity})), the green solid curves in (a2) and (b2) indicate the analytical results of the winding number (see Methods), the vertical dashed lines indicate the locations of topological phase transitions from theoretical predictions, and the other curves indicate the numerical simulation results. The coin parameters are the same as those of Fig.~\ref{moments} in the main text.}
\label{fig:suppfig1}
\end{figure}

For the non-unitary QWs in general, analytic expressions such as Eq.~(\ref{eqn:groupvelocity}) are typically unavailable. From numerical calculations (see Fig.~\ref{fig:suppfig1}), we see that signatures of topological phase transitions in the second moments persist in the non-unitary cases. In fact, at short time steps or away from the topological phase boundary, the second moments from the unitary and the non-unitary QWs are almost the same. However, at longer time steps, precipitous dips emerge in the second moment of non-unitary QWs near topological phase transitions. Such a behavior can be explained by mapping the Floquet operators in Eqs.~(1), (3) and (4) to operators with the so-called pseudo-unitarity.

For such a purpose, we replace $M$ with $\gamma M$ in Eqs.~(1), (3) and (4) of the main text, and define $\overline{U}_i':=\gamma\widetilde{U}_i'$ and $\overline{U}_i'':=\gamma \widetilde{U}_i''$, with $i=3,4$ and $\gamma=(1-p)^{-\frac{1}{4}}$. Following the definition of winding numbers in the previous sections, it is straightforward to show that the topological phase diagrams for QWs are not changed with the introduction of $\gamma$ in the Floquet operators. Further, as $\gamma$ is a constant, it only introduces a spatially homogeneous decay $\gamma^{t}$ to the walker at the $t$-th step, which does not change the statistical moments at any given time. Most importantly, as we will show in the next section, both $\overline{U}_i'$ and $\overline{U}_i''$ have pseudo-unitaritary regions on the phase diagram, which depend on both the loss parameter $p$ and the coin parameters.

In Figs.~\ref{fig:suppfig2}(a) and~\ref{fig:suppfig2}(b), we show the boundary between regions with pseudo-unitarity and those without using red lines. Typically, the pseudo-unitarity is lost in regions surrounding the topological phase boundaries. As pseudo-unitarity is a necessary and sufficient condition for the reality of the quasienergy spectrum of the effective non-Hermitian Hamiltonian, the loss of pseudo-unitarity leads to imaginary-valued quasienergies at certain points in momentum space. The resultant non-pseudo-unitary QW has similar behaviour to a non-unitary QW with a broken parity-time symmetry, in that the long-time spatial distribution of the walker is Gaussian-like rather than ballistic (see Fig.~\ref{fig:suppfig2}(c)). Hence, the second moment decreases rapidly close to a topological phase transition, which carries over to the quantum-walk dynamics governed by the operators $\widetilde{U}_i'$ and $\widetilde{U}_i''$, so long as the evolution time is long enough.

\begin{figure}
\includegraphics[width=0.8\textwidth]{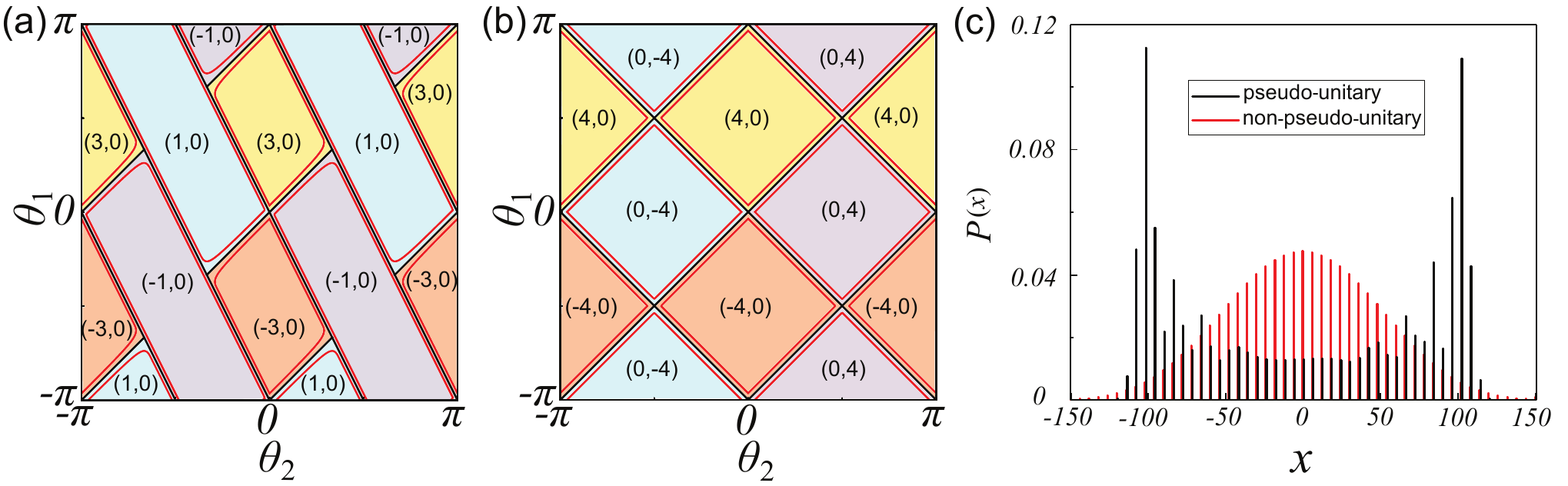}
\caption{Phase diagrams of three-step and (b) four-step Floquet operators $\overline{U}_i'$ and $\overline{U}_i''$ ($i=3,4$). Different topological phases are characterized by the winding numbers $(\nu',\nu'')$. The black lines mark the topological phase boundaries, which are the same as those of $\widetilde{U}_i'$ and $\widetilde{U}_i''$. The red lines represent the boundaries between regions with pseudo-unitarity and those without. The pseudo-unitarity is lost in a loss-dependent region around the topological phase boundary. As the loss parameter $p$ increases, the non-pseudo-unitary region also increases. (c) The long-time ($t=50$) spatial distributions of the walker governed by the Floquet operator $\overline{U}_3'$ in the pseudo-unitary (black) and the non-pseudo-unitary (red) regions. The coin parameters are ($\theta_1=\pi/4,\theta_2=0$) and ($\theta_1=0,\theta_2=0$), respectively, for the pseudo-unitary and the non-pseudo-unitary case. Note that the spatial distributions are the same as the normalized spatial distributions under $\widetilde{U}_3'$ with the same coin parameters.}
\label{fig:suppfig2}
\end{figure}

\section{Pseudo-Unitarity}
In this section, we define and discuss pseudo-unitarity. We show that pseudo-unitarity of a Floquet operator $U$ is equivalent to the reality of the quasienergy spectrum of the corresponding effective Hamiltonian~\cite{M2002a,M2002b,M2004}.
A necessary and sufficient condition for the spectrum of a non-Hermitian Hamiltonian to be purely real can be formulated in terms of pseudo-Hermiticity~\cite{M2002a,M2002b}. Such a condition can be generalized to the Floquet operator, where a Floquet operator $U$ has $\eta$-pseudo-unitarity~\cite{M2004} if it satisfies $U^{-1}=\eta U^\dag\eta^{-1}$; here $\eta$ is a Hermitian invertible linear operator.

In general, a non-unitary Floquet operator $U$ has a complete set of biorthonormal eigen-vectors $\{|\psi_{\pm}\rangle,|\chi_{\pm}\rangle\}$. Therefore, in momentum space,
\begin{equation}
\begin{split}
U_k = &~ n_0\sigma_0-\text{i}n_1\sigma_x-\text{i}n_2\sigma_y-\text{i}n_3\sigma_z,~
U_k^\dag = n_0\sigma_0+\text{i}n^*_1\sigma_x+\text{i}n^*_2\sigma_y+\text{i}n^*_3\sigma_z,\\
|\psi_{\pm}\rangle=&~\frac{1}{\sqrt{2\sqrt{1-n^2_0}(\sqrt{1-n^2_0}\pm n_3)}}\left(n_3\pm\sqrt{1-n^2_0},n_1+\text{i}n_2\right)^T,\\
\langle\chi_{\pm}|=&~\frac{1}{\sqrt{2\sqrt{1-n^2_0}(\sqrt{1-n^2_0}\pm n_3)}}\left(n_3\pm\sqrt{1-n^2_0},n_1-\text{i}n_2\right),\\
U_k|\psi_\pm\rangle=&~\lambda_\pm|\psi_\pm\rangle,~U^\dag|\chi_\pm\rangle=\lambda^*_\pm|\chi_\pm\rangle,
~\langle\chi_\mu|\psi_\nu\rangle=\delta_{\mu\nu},~\sum_\mu|\psi_\mu\rangle\langle\chi_\mu|=1,
~U_k=\sum_\mu\lambda_\mu|\psi_\mu\rangle\langle\chi_\mu|,
\end{split}
\end{equation}
where $\lambda_\pm=n_0\mp \text{i}\sqrt{1-n^2_0}$. Note the parameters $\lambda_{\pm}$, $n_0$,$n_1$,$n_2$, and $n_3$ are all momentum-dependent.

We define the effective Hamiltonian via $U_k=\exp (-\text{i} H_{\rm eff})$. The quasienergy of $H_{\rm eff}$ is real if and only if $|\lambda_\pm|=1$, which is the case when $n_0^2\leq 1$. Let $\{|\phi_{\mu}\rangle\}$ be an arbitrary complete orthonormal basis, i.e., $\langle \phi_{\mu}|\phi_{\nu}\rangle=\delta_{\mu\nu},~\sum_{\mu=\pm}|\phi_{\mu}\rangle\langle\phi_{\mu}|=1$ (for example $|\phi_+\rangle=|+\rangle,~|\phi_-\rangle=|-\rangle$). We define $O:=\sum_\mu|\psi_\mu\rangle\langle \phi_\mu|$ and $U_0:=\sum_\mu\lambda_\mu|\phi_\mu\rangle\langle \phi_\mu|$. It is straightforward to show that $O$ is invertible with the inverse given by $O^{-1}=\sum_\mu|\phi_\mu\rangle\langle\chi_\mu|$, and $O^{-1}U_kO=U_0$. While $|\lambda_\pm|=1$, $U_0$ is unitary with $U_0U^\dag_0=1$. Therefore, we have $O^{-1}U_kO(O^{-1}U_kO)^\dag=1$. Defining $\eta:=OO^\dag$, we have $U_k^{-1}=\eta U_k^\dag(\eta)^{-1}$; i.e., $U_k$ is $\eta$-pseudo-unitary. Pseudo-unitarity of $U_k$ is the direct result of the reality of the quasienergy at the momentum $k$.

Conversely, if $U_k$ is $\eta$-pseudo-unitary, we have $U_0U^\dag_0=1$, which leads to $|\lambda_\pm|=1$ and the reality of the quasienergy of $H_{\rm eff}$ at the corresponding momentum $k$. Thus the reality of the quasienergy is equivalent to the pseudo-unitarity of the Floquet operator $U_k$. It is also apparent that the pseudo-unitarity of $U_k$ breaks down when $n^2_0> 1$.

The Floquet operators $\overline{U}_i'$ and $\overline{U}_i''$ defined in the previous section possess pseudo-unitarity when their corresponding $n^2_0\leq 1$ for all $k$. The boundaries between regions with pseudo-unitarity and those without are therefore calculated by requiring $n_0^2=1$ be satisfied for at least one $k$. We plot the boundary in Fig.~\ref{fig:suppfig2} in red. It appears that, in both cases, the pseudo-unitarity is lost in the immediate vicinities of topological phase boundaries. We note that, as $p$ increases, the widths of the non-pseudo-unitary regions also increase.

\section{Average chiral displacement}
In this section, we confirm topological phase transitions in our quantum-walk dynamics by measuring the average chiral displacement~\cite{Cardano2017}
\begin{equation}
C(t)=\frac{\sum_x\langle \psi_t|x\rangle\langle x|\otimes \Gamma|\psi_t\rangle}{\sum_x\langle\psi_t|x\rangle\langle x|\otimes\one_\text{c}|\psi_t\rangle}.
\end{equation}
To measure the average chiral displacement, a HWP with setting angle $22.5^\circ$ and a polarizing beamsplitter (PBS) are inserted between the last sandwich-type HWP-PPBS-HWP setup and avalanche photodiodes (APDs). Here, PPBS is the abbreviation for a partially polarizing beamsplitter. The HWP applies a basis transformation on the polarizations of photons which have been transmitted by the sandwich-type setup, and the following PBS projects the photons into the basis states $\{\ket{+},\ket{-}\}$. The average chiral displacement is then
\begin{equation}
C_\text{exp}(t)=\frac{\sum_x x\left[N_\text{T}(x,t)- N_\text{R}(x,t)\right]}{\sum_{x'}\left[N_\text{T}(x',t)+N_\text{R}(x',t)\right]}.
\end{equation}

Surprisingly, while the chiral symmetry is broken in our non-unitary QWs, the scaled chiral displacement $-2C(t)$ still oscillates around the integer-valued winding numbers (see Fig.~\ref{fig:suppfig3}). At the topological phase boundaries, the scaled chiral displacements feature large jumps, whose locations are consistent with topological phase boundaries measured from loss and from the statistical moment. We note that signals in the average chiral displacement should be improved for QWs with large time steps.

\begin{figure}
\includegraphics[width=0.8\textwidth]{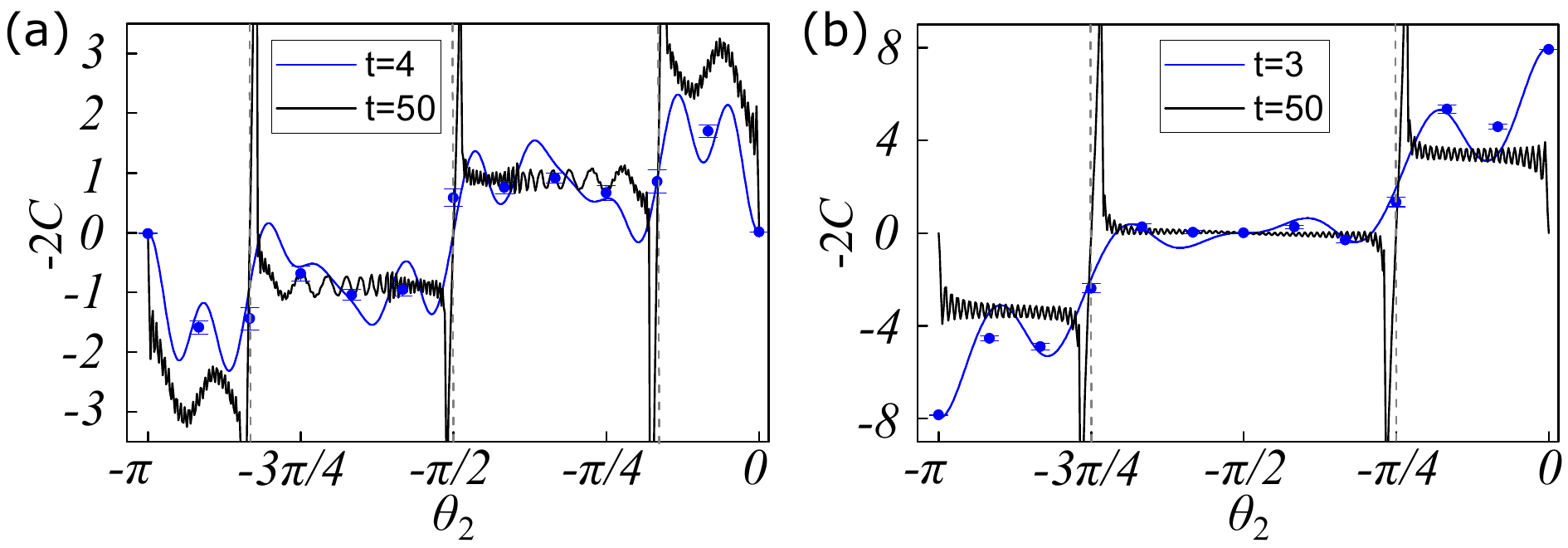}
\caption{Measured scaled chiral displacements $-2C$ of the walker position distribution for three-step non-unitary QWs governed by $\widetilde{U}_3'$ up to $4$ time steps (upper layer); and for four-step non-unitary QWs governed by $\widetilde{U}_4'$ up to $3$ times steps (lower layer). The coin parameters $(\theta_1,\theta_2)$ are scanned along the dotted lines in the phase diagrams of Figs.~\ref{displacement3}(a) and \ref{displacement4}(a). The loss parameter is fixed at $p=9/25$. The solid curves indicate the numerical simulations for QWs up to $4$ (or $3$) steps and $50$ steps and the experimental results are presented by dots. The vertical dashed lines indicate locations of topological phase transition from theoretical predictions. Experimental errors are due to
photon counting statistics. 
}
\label{fig:suppfig3}
\end{figure}

\section{Topological edge states}
In this section, we confirm the topological properties of non-unitary FTPs with large winding numbers by the experimental observation of localized edge states at boundaries between regions with different winding numbers.

To probe edge states, we implement inhomogeneous QWs with a fixed loss parameter $p=9/25$. We fixe coin parameters for the left region ($x<0$) and vary those of the right region ($x\geq 0$), such that a boundary is created near $x=0$. As shown in Fig.~\ref{fig:newfig5}(a,c), for both the three- and four-step non-unitary QWs, there are no localized edge states when the left and right regions belong to the same FTP characterized by the same set of winding numbers. In contrast, localized edge states emerge when the left and right regions feature distinct winding numbers, as shown in Fig.~\ref{fig:newfig5}(b,d).

Here, the corrected probability is defined as $P_\text{C}(x,t)=\gamma^{2t} \bra{\psi_{t-1}}\widetilde{U}'^{\dagger} \left(\ket{x}\bra{x}\otimes\one_\text{c}\right) \widetilde{U}'\ket{\psi_{t-1}}$, where $\gamma=(1-p)^{-\frac{1}{4}}$. Experimentally,  the corrected probability can be probed by photon counts
of the transmitted photons after step $t$ at the position $x$ via a coincidence measurement to the total number of transmitted photons, i.e., $\gamma^{2t} \frac{N_\text{T}(x,t')}{\sum_{x'}\left[\sum^{t}_{t''=1}N_\text{R}(x',t'')+N_\text{T}(x',t)\right]}$.

\begin{figure}
\includegraphics[width=\textwidth]{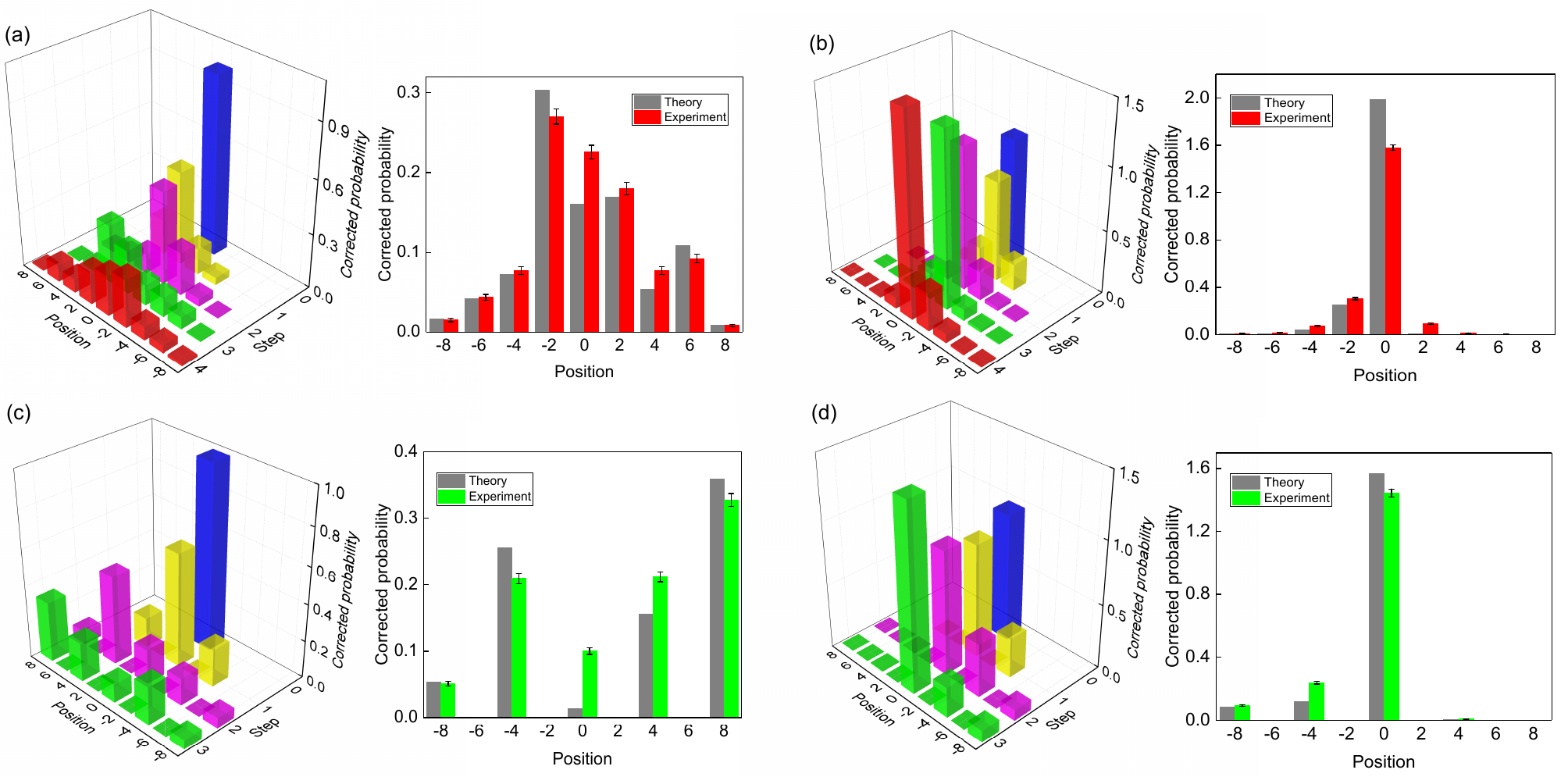}
\caption{Experimental observation of topological edge states in inhomogeneous QWs with a fixed loss parameter $p=9/25$. (a)-(b) Measured corrected probability distribution of three-step non-unitary QWs governed by $\widetilde{U}'_3$ with fixed coin parameters for the left region $(\theta_1^l,\theta_2^l)=(2\pi/3,\pi/4)$, corresponding to $(\nu',\nu'')=(1,0)$; and with varying coin parameters for the right region $(\theta_1^r,\theta_2^r)=(3\pi/4,\pi/4)$ (a) and $(\theta_1^r,\theta_2^r)=(-9\pi/10,3\pi/5)$ (b), corresponding to $(\nu',\nu'')=(1,0)$ and $(\nu',\nu'')=(-1,0)$, respectively.
(c)-(d) Measured corrected probability distribution of three-step non-unitary QWs governed by $\widetilde{U}'_4$ with fixed coin parameters for the left region $(\theta_1^l,\theta_2^l)=(\pi/16,5\pi/16)$, corresponding to $(\nu',\nu'')=(0,4)$; and with varying coin parameters for the right regions $(\theta_1^r,\theta_2^r)=(-\pi/8,3\pi/4)$ (c) and $(\theta_1^r,\theta_2^r)=(-9\pi/16,-5\pi/16)$ (d), corresponding to $(\nu',\nu'')=(0,4)$ and $(\nu',\nu'')=(-4,0)$, respectively.
Left column: measured corrected probability distributions up to $4$ time steps ($3$ time steps) . Right column: comparison between the measured and numerically calculated corrected probability distribution at the last step.}
\label{fig:newfig5}
\end{figure}

\section{Robustness against disorder}
We now check the robustness of topological properties of our system against small perturbations.
We find that the quantization of the average displacement of the multi-step non-unitary QW here is robust against both static and dynamic disorders. Our results therefore not only confirm the robustness of the measurement scheme, but also demonstrate the robustness of the FTPs with large topological invariants. Here, we use the three-step non-unitary QW for the evolution operator $\widetilde{U}_3'$ with loss parameter $p=1$ as an example.

First, we test the robustness of the quantization of the average displacement against static disorder. We keep the mean values of the coin parameters $\langle \theta_1\rangle$ and $\langle\theta_1\rangle$ on the line $\langle\theta_1\rangle=\langle\theta_2\rangle+\pi/2$ and measure the probabilities of the three-step non-unitary QW up to $4$ time steps. We implement quantum-walk dynamics governed by the evolution operator $\widetilde{U}_3'$ with $10$ randomly generated coin rotations $R(\langle\theta_{1,2}\rangle+\delta\theta)$ for each position. For static disorder, the time-independent $\delta\theta$ is unique for each position and chosen from the intervals $\left[-\pi/20,\pi/20\right]$. In our experiment, $\delta\theta$ is implemented by manipulating the setting angles of HWPs by small random amounts $\delta\theta$ around the coin parameters $(\theta_1,\theta_2)$. We show in Fig.~\ref{disorder1}(a) mean values of $10$ sets of average displacements, which are still quantized, as expected.

Second, we study the effect of the dynamic disorder. To generate dynamic disorder, a time-dependent coin rotation is required. The setting angles of HWPs for each step are modulated by a small random amount around the coin parameters $(\theta_1,\theta_2)$. The strength of the disorder is determined by the angle shift $\delta\theta$, which is randomly generated at each time step from the interval $\left[-\pi/20,\pi/20\right]$. Note that $\delta\theta$ here is time-dependent but spatially homogeneous. We measure the probabilities and calculate the mean values of the $10$ sets of average displacements. The results shown in Fig.~\ref{disorder1}(b) agree with theoretical predictions.

\begin{figure}
\includegraphics[width=0.8\textwidth]{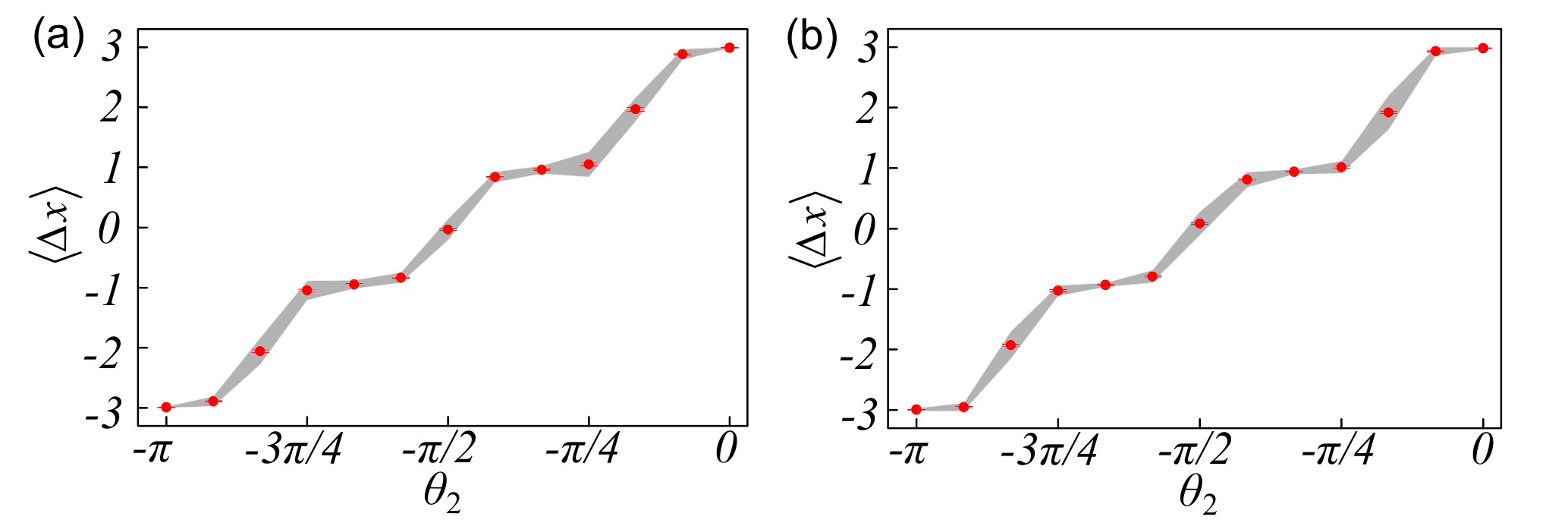}
\caption{ Average displacements for three-step non-unitary QWs governed by $\widetilde{U}_3'$ with either static disordered rotation angles (a) or dynamic disordered rotation angles (b). The loss parameter is fixed at $p=1$. The disordered rotation angles are given by $\theta_{1,2}+\delta\theta$, where $\delta\theta$ is chosen from the interval $\left[-\pi/20,\pi/20\right]$. For static disorder, $\delta\theta$ is unique for each position and is independent of time. For dynamic disorder, $\delta\theta$ is unique for each time step and is independent of the position of the walker. The coin parameters $(\theta_1,\theta_2)$ are scanned along the dotted line in the phase diagram (Fig.~\ref{displacement3}(a) in the main text). The symbols and the grey shadings, respectively, indicate mean values of the measured average displacements and the range of the standard deviations averaged over $10$ different ensembles for each pair of $(\theta_1,\theta_2)$. Experimental errors are due to
photon-counting statistics.}
\label{disorder1}
\end{figure}

Finally, we confirm the robustness of topological edge states against static disorder. As shown in Fig.~\ref{disorder2}, localized topological edge states persist in the presence of static disorder for both the three- and four-step non-unitary QWs.

\begin{figure}
\includegraphics[width=0.8\textwidth]{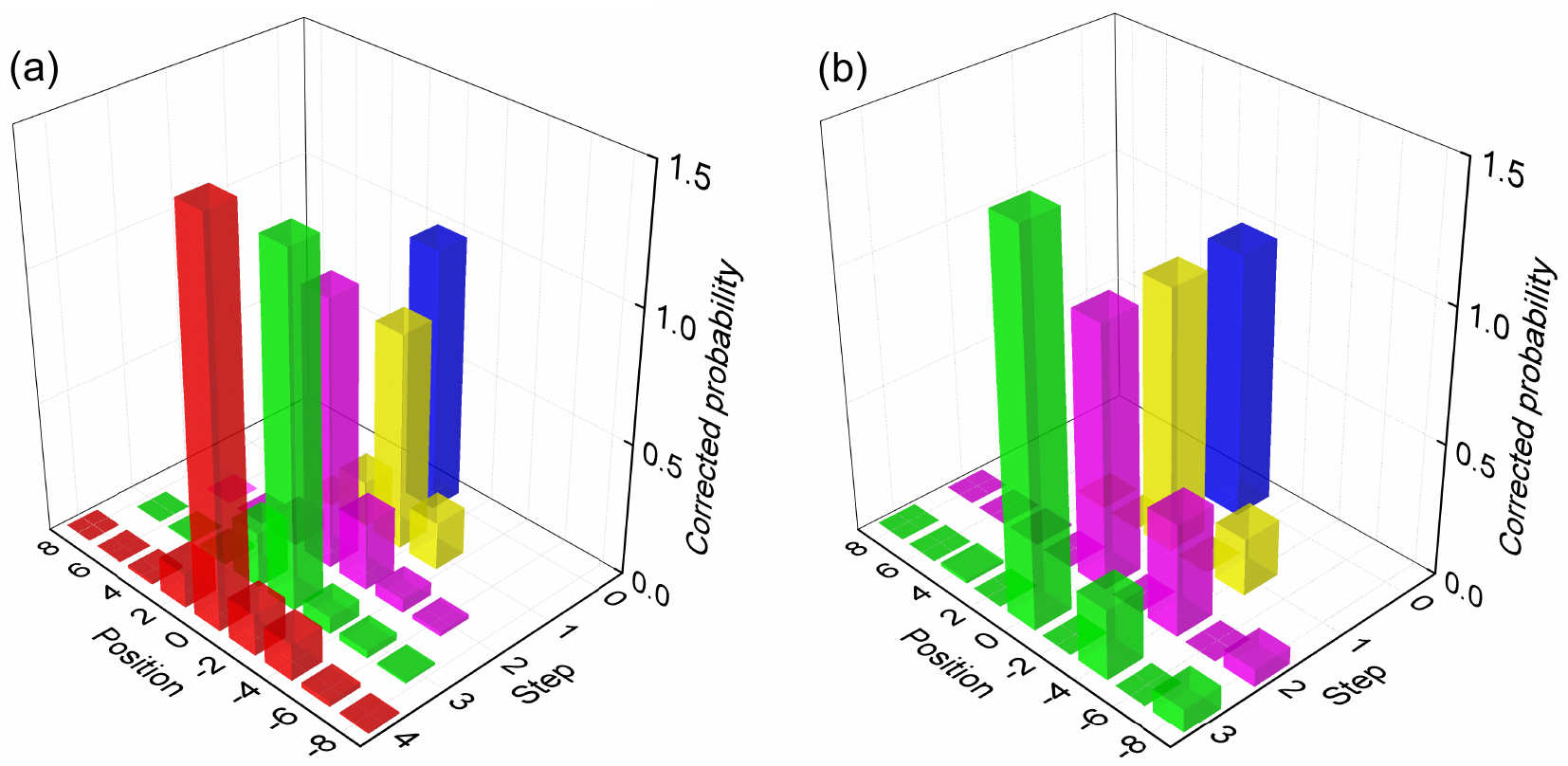}
\caption{Robustness of edge states against static disorder. (a) Probability distributions of QWs governed by $\widetilde{U}'_3$ up to $4$ time steps. The coin parameters are the same as those in Fig.~\ref{fig:newfig5}(b). (b) Probability distributions of QWs governed by $\widetilde{U}'_4$ up to $3$ time steps. The coin parameters are the same as those in Fig.~\ref{fig:newfig5}(d). The rotation angles are given by $\theta_{1,2}+\delta\theta$, where the time-independent $\delta\theta$ is unique for each position and is chosen from the interval $\left[-\pi/20,\pi/20\right]$.}
\label{disorder2}
\end{figure}

\end{widetext}

\end{document}